\documentclass{aa}

\usepackage[usenames,dvipsnames,svgnames,table]{xcolor}
\usepackage{float}
\usepackage[
breaklinks=true,
hyperindex=true,
colorlinks=true,
linkcolor=Sepia,
citecolor=Purple]{hyperref}
\usepackage{bm}
\usepackage{multirow}
\usepackage{booktabs}
\usepackage{arydshln}

\usepackage{graphicx}

\makeatletter
\renewcommand*\aa@pageof{, page \thepage{} of \pageref*{LastPage}}
\makeatother

\usepackage[varg]{txfonts}

\usepackage{natbib}
\bibpunct{(}{)}{;}{a}{}{,} 

\definecolor{ogreen}{rgb}{0.07,0.54,0.03}

\defcitealias{brucyInefficientStarFormation2024}{Paper II}

\begin{document}

\title{Inefficient star formation in high Mach number environments
}
\subtitle{I. The turbulent support analytical model} 

\titlerunning{The inefficient high Mach regime for star formation. I. The turbulent support analytical model}

   \author{Patrick Hennebelle  \inst{\ref{inst1}} 
   \and  Noé Brucy \inst{\ref{inst2}} 
   \and  Tine Colman \inst{\ref{inst1}} 
    }

   \institute{ 
    Université Paris-Saclay, Université Paris Cité, CEA, CNRS, AIM, 91191, Gif-sur-Yvette, France,
      \label{inst1} 
      \and 
      Universität Heidelberg, Zentrum für Astronomie, Institut für Theoretische Astrophysik, Albert-Ueberle-Str 2, D-69120 Heidelberg, Germany,
      \label{inst2}
}

\date{Received \today; submitted to A\&A}

\abstract
{The star formation rate (SFR), the number of stars formed per unit of time, 
is a fundamental quantity in the evolution of the Universe.}
{While turbulence is believed to play a crucial role in setting the SFR, the 
exact mechanism remains unclear. Turbulence promotes star formation by compressing 
the gas, but also slows it down by stabilizing the gas against gravity.
Most widely used analytical models rely on questionable assumptions, including: 
$i)$ integrating over the density PDF, a one-point statistical description 
that ignores spatial correlation, $ii)$ selecting self-gravitating gas based 
on a density threshold that often ignores turbulent dispersion, 
$iii)$ assuming the freefall time as the timescale for estimating SFR 
without considering the need to rejuvenate the density PDF, 
$iv)$ assuming the density probability distribution function (PDF) to be log-normal. This leads to the reliance 
on fudge factors for rough agreement with simulations. Even more seriously, when a more accurate density PDF is being used, the  classical theory
predicts a SFR that is essentially 0.}
{Improving upon the only existing model that incorporates the spatial 
correlation of the density field, we present a new analytical model that, 
in a companion paper, is rigorously compared against a large series of 
numerical simulations. We calculate the time needed to rejuvenate density 
fluctuations of a given density and spatial scale, revealing that it is 
generally much longer than the freefall time, rendering the latter 
inappropriate for use.}
{We make specific predictions regarding the role of the Mach number, 
${\cal M}$, and the driving scale of turbulence divided by the mean Jeans 
length. At low to moderate Mach numbers, turbulence does not reduce and may 
even slightly promote star formation by broadening the PDF. However, at 
higher Mach numbers, most density fluctuations are stabilized by turbulent 
dispersion, leading to a steep drop in the SFR as the Mach number increases. 
A fundamental parameter is the exponent of the power spectrum of the natural logarithm of the density, $\ln \rho$, 
characterizing the spatial distribution of the density field. In the high 
Mach regime, the SFR strongly depends on it, as lower values imply a paucity 
of massive, gravitationally unstable clumps.}
{We provide a revised analytical model to calculate the SFR of a system, 
considering not only the mean density and Mach number but also the spatial 
distribution of the gas through the power spectrum of $\ln \rho$, as well 
as the injection scale of turbulence. At low Mach numbers, the model predicts 
a relatively high SFR nearly independent of ${\cal M}$, whereas for high 
Mach, the SFR is a steeply decreasing function of ${\cal M}$.}

\keywords{stars: formation --- ISM: clouds --- physical processes: turbulence}

\maketitle
\section{Introduction}

One of the major goals of modern astrophysics is to understand how and when structures formed in the Universe. 
In particular, the speed at which a gaseous system, like a molecular cloud, undergoes gravitational collapse and produces 
stars is of great importance. Estimating the star formation rate (SFR) in various systems has been the subject of 
many observational studies during the last decades \citep[e.g.][]{schmidt1959,lada2010,kennicutt2012,schrubaHowGalacticEnvironment2019,sunStarFormationLaws2023} and remains an active field of research as new telescopes such as JWST push the boundaries of extra-galactic observations and resolve star forming regions within the Milky Way with extraordinary detail.
A major puzzle is the remarkably small value of the SFR, typically found to be about two orders of magnitude below the 
simplest estimate obtained by dividing the mass of dense gas by the freefall time of the system. This fact is generally 
interpreted as the consequence of support against gravitational collapse, which could be due to either magnetic fields 
\citep{shu1987} or turbulence \citep{zuckermanRadioRadiationInterstellar1974,zuckermanModelsMassiveMolecular1974,maclow2004}. The role of stellar feedback is also heavily emphasised, particularly 
through numerous numerical simulations \citep[e.g.][]{kim2013,walch15,iffrig2017}. In this context, the exact 
distinction between the role played by feedback and turbulence is partially unclear. Feedback is certainly a source of 
turbulence, but it also creates diverging motion that destroys star-forming clouds, limiting the star formation efficiency of these clouds.

In an attempt to understand and predict how the SFR is determined, several models have been developed, particularly during 
the last two decades \citep{krum2005,padoan2011,hc2011,renaud2012}. Most of these models, as presented in more detail below, rely on the 
idea that only gas above a certain density is undergoing collapse. Therefore, they integrate the density probability 
density function (PDF) above a selected density threshold that generally derives from the sonic and the Jeans length. A somewhat 
different approach has been followed by \citet{hc2011}, where it is proposed to proceed in two steps. First, the mass 
spectrum of self-gravitating clouds is inferred. This step entails the spatial coherence, that is the power spectrum, of the 
density field (more precisely, the natural logarithm of the density field). Once this step is completed, the sum over this mass spectrum divided by the freefall time provides an estimate for the SFR. Importantly, the density threshold used in this approach
is not unique but rather is scale dependent. 
More over, what limits the integration is the maximum size of a density 
fluctuation that may exist in the system.

Recently, \citet{brucy2020,brucy2023} have performed a series of calculations in which both stellar feedback 
and large-scale turbulent driving are operating and responsible for the driving of turbulence. It has been shown that 
sufficiently vigorous driving could substantially reduce the star formation rate in these simulations. Indeed, provided the 
driving was strong enough, the Schmidt-Kennicutt relation \citep{kennicutt2012} could be reproduced. The observed 
behaviour of a very severe SFR decrease is not obviously compatible with the existing models. In a companion paper \citep[hereafter \citetalias{brucyInefficientStarFormation2024}]{brucyInefficientStarFormation2024}, we 
thoroughly investigate this issue and find that the existing models present poor agreement with a series of 
simulation results. 

An accurate model for the SFR is not only important for the understanding of the precise physical origin of the SFR.
It also has practical uses.
In many, if not all simulations of galaxies, the numerical resolution is not sufficient to describe the 
star-forming gas well enough to obtain a realistic estimate of the SFR, and a subgrid model needs to be used. This is, for 
instance, the case in \citet{braun2015} and \citet{nunez2021}.

In this paper, we start with a review of the existing analytical models in Section two. 
We discuss the various steps and assumptions that one has to make, after which we argue that some improvements are necessary. Two of them, namely the density 
PDF and the control time of star formation are presented and 
discussed in Section three and four. 
In Section five, we incorporate these improvements into the derivation of a revised analytical model and we examine the consequences
they have regarding the self-gravitating clump mass spectrum.
Then in Section six we study the resulting SFR. 
This leads us to identify the existence of two regimes. In particular, 
we discover a new regime of inefficient star formation at high Mach numbers.
We determine the threshold Mach number, which separates the low and high Mach number regimes and discuss why the previous models did not recover this behaviour.
Section seven explores the influence of the model parameters on the resulting Mach number -- SFR relation.
In \citetalias{brucyInefficientStarFormation2024}, we compare the new model presented in this work to the results of a suite of numerical simulations. Excellent agreement is obtained. In the eighth session, we discuss our 
finding in the context of a galaxy and discuss various possible improvements. Section nine concludes the paper.

\section{Overview of existing star formation theories}
\label{sec:existing_models}
In this section, we review in some details the relevant physical processes and the various assumptions used in present analytical models that infer the 
SFR. 

\begin{table}[ht]
	\caption{Main notations and abbreviations used in the article.}
  \begin{center}
	\begin{tabular}{lp{6cm}}
	\toprule
	Notation &  Description \\
    \midrule
 	$G$                       & Gravitational constant \\
    $m_p$                     & Mass of a proton \\
    $\mu = 1.4$               & Mean molecular weight, in unit of $m_p$ \\
    \midrule
    $M_0$                     & Total mass in the system\\
    $L_0$                    & Scale of the system \\
    $R_0 = L_0/2$             & Typical radius of the system \\
    $n_0$                     & Mean number density\\
    $\rho_0 = n_0 \mu\ m_p$    & Mean density \\
    $\sigma_0$                & Velocity dispersion at the system scale $L_0$\\
    $c_\mathrm{s}$            & Sound speed \\
    $L_\mathrm{i}$             & Injection length of turbulence \\
    $y_\mathrm{cut}$                 &  largest radius of structures considered is  $y_\mathrm{cut} L_\mathrm{i} $  \\
    \midrule
    $\sigma$                  & 3D mass-weighted velocity dispersion\\
    $\mathcal{M}$             & 3D Mach number  \\
    $\mathcal{M}_\mathrm{thres}$ & Mach number at which the transition between the efficient and inefficient star formation regimes occurs \\
    $\mathcal{M}_\mathrm{eq}$ & Expected Mach number in galaxies from dynamical equilibrium\\
    $\tau_\mathrm{ff}$        & Free-fall time \\
    $\tau_\mathrm{rep}$        & Replenishment time \\
    $\tau_\mathrm{cont}$ 
    & Control time for star formation\\
    $\mathrm{SFR}_\mathrm{ff}$  &  Dimensionless SFR per free-fall time \\
    \midrule
    $\delta = \ln (\rho / \rho_0) $ & Logarithmic density contrast \\
    $S_\delta$             & Volume weighted variance of $\delta$\\
    $T_{\mathrm{CH}}$         & Temperature of turbulence  \citep[][]{castaing1996} \\
    \midrule
    $M_J$                     & Jeans mass \\
    $\lambda_J$               & Jeans length \\
     $\lambda_\mathrm{turb} = \mathcal{M}\lambda_J $     & Turbulent Jeans length \\
    $\widetilde{R} = R / \lambda_J$ & For any spatial quantity $R$, $\widetilde{R}$ is the same quantity divided by the Jeans length. \\
    $\widetilde{M} = M / M_J$ & For any mass $M$, $\widetilde{M}$ is the same quantity divided by the Jeans mass. \\
    \midrule
    $n_v$ & Exponent of velocity powerspectrum \\
    $\eta_v$ & $\eta_v = (n_v-3) / 2$ \\
    $n_d$ & Exponent of the natural logarithm of density  \\ & powerspectrum \\
    $\eta_d$ & $\eta_d = (n_d-3) / 2$ \\
	\bottomrule
	\end{tabular} 
  \end{center}
	\label{tbl:notation}
\end{table}

\subsection{Definitions and concepts}

Before reviewing the theories from the literature, we
give an overview of important definitions and recurring elements used in analytical star formation theories.
This also serves as a guide to familiarise the reader with the notations used in this work, which are summarised in Table~\ref{tbl:notation}.

\subsubsection{Characterisation of a star-forming system}
First, we define the system over which the SFR is measured. 
In practice, this could be an entire galaxy, a sub-region of a galaxy, an individual star forming cloud...
To aid the interpretation, we will assume the system to be a hypothetical cloud of mass $M_0$, radius $R_0$
and mean density $\rho_\mathrm{0}$. These properties are related through
\begin{equation}
\label{eq:volume}
    \rho_0 = \frac{M_0}{a_\mathrm{g} R_0^3},
\end{equation}
where $a_\mathrm{g}$ represents a general factor used in the calculation of the volume of the system, which varies with its geometry: $a_\mathrm{g}=1$ for a cube with side $R_0$ and $a_\mathrm{g}=4 \pi/ 3$ for a sphere of radius $R_0$.

The interstellar mediums (ISM) is found to be highly turbulent.
We denote the velocity dispersion measured on the scale of the entire system with $\sigma_0$.
We assume the cloud is isothermal with a constant sound speed $c_\mathrm{s}$, which is a reasonable approximation for cold ISM clouds.
The Mach number, measured on the scale of the cloud, is then simply $\mathcal{M}_0 = \sigma_0 / c_\mathrm{s}$.

\subsubsection{The log-normal density PDF}

A recurring ingredient in present-day star formation theories is the density PDF. This one-point statistic describes how much gas exists at each density within the system.
It is usually expressed in terms of the logarithmic density contrast
\begin{equation}
\nonumber
    \delta = \ln(\rho/ \rho_0),
\end{equation}
which is dimensionless.
Several simulations \citep[e.g.][]{kritsuk2007,fed2013} have found that supersonic turbulence leads to density PDFs that are well-represented by a log-normal form \citep{vazquez1994,nordlund1999,Federrath08}:
%
%
\begin{eqnarray}
\label{Pr0}
{\cal P}(\delta) &=& \dfrac{1}{\sqrt{2 \pi S_\delta}} 
\exp\left(- \dfrac{(\delta - \bar{\delta})^2}{2 S_\delta}\right) \; \mathrm{with} \\
\label{eq:S_from_mach}
S_\delta &=& \ln\left(1 + b^2 {\cal M}^2\right), \text{ and }  \\
\bar{\delta} &=& -S_\delta/2.\nonumber
\end{eqnarray}
The Mach number $\mathcal{M}$ controls the width of the density PDF through the variance $S_\delta$: the more turbulent the medium, the more high density fluctuations are created.
The parameter $b$ is set by the nature of the turbulence driving, which can be either purely compressive ($b \simeq 1$), purely solenoidal ($b \simeq 0.3$) or a mix of both. The interstellar medium usually contains mixed modes with $b \simeq 0.67$.
Alternative forms for the density PDF exist and may be more appropriate than the log-normal PDF in certain situations, as will be discussed later in this work.

\subsubsection{Spatial correlations from turbulence}

A characteristic of turbulence is that it introduces motion and structure on a wide ranges of scales. 
Turbulence generates velocity and density power spectra which follow power-laws.

The velocity dispersion is then given by 
\begin{equation}
\label{eq_turb}
    \sigma = \sigma_0 \left( {\frac{R}{R_0}} \right) ^{\eta_v},
\end{equation}
with $\eta_v = 0.3-0.5$.
Such a relation is indeed recovered in observations, where it is known as the Larson relation
\citep{Larson81,HF12}. Classically, $\eta_v \approx 0.5$, but variations have been found between studies.
The sonic length $\lambda_\mathrm{s}$ is the scale at which the gas become subsonic, that is $\sigma(\lambda_\mathrm{s}) = c_\mathrm{s}$. 
A useful definition is $n_v=3+2 \eta_v$ in which case 
$\widetilde{\sigma} ^2 \propto k^{-n_v}$, where $\widetilde{\sigma}^2$
is the powerspectrum of $\sigma$. In incompressible turbulence $n_v= 11/3$.


Another quantity that will be important below is the 
powerspectrum of $\ln \rho$.
Whereas this latter  has not been systematically 
investigated, several studies have calculated the powerspectrum of the density, 
$\rho$ \citep[e.g.][]{kim2005,konstandin2016}. It has been found that as the Mach number increases, the density powerspectrum becomes gradually flatter.
Due to the strong shocks, the density distribution develops 
sharp contrast. For instance \citet{konstandin2016}, performing a series of simulations, have proposed that the density powerspectrum exponent follows 
$-2 -2.1 \mathcal{M}^{-0.33}$.
In particular, in the limit of very strong Mach numbers, this exponent is going to -2, a value predicted by \citet{tatsumi1974}.
Since $\ln \rho$ presents much less prominent contrasts than $\rho$, one 
expects that the exponent of the powerspectrum of $\ln \rho$, which we denote as $n_d$ in this work, presents less 
pronounced stiffening at high Mach numbers. 
As for the velocity power spectrum, it is convenient to define the quantity  $\eta_d$ such as
\begin{equation}
\label{eq:eta_d}
    \eta_d = (n_d-3)/2
\end{equation}

In the companion paper, a 
systematic investigation is presented, and it is found that at high Mach 
numbers, the exponent is approaching~$\simeq -3$, whereas at low Mach numbers, it is about~$\simeq -3.8$. 
This corresponds to values of 
$\eta_d \simeq 0$ at high Mach number to $\eta_d \simeq 0.4$ at low Mach numbers.

\subsubsection{Gravitational collapse}

Gravity plays a central role in the formation of stars.
The virial theorem quantifies whether a structure is stable, collapsing or expanding: 
\begin{equation}
    \frac{1}{2} \frac{\mathrm{d}^2 I}{\mathrm{d} I^2} \approx
    2 E_\mathrm{kin} + 2 E_\mathrm{therm} + E_\mathrm{pot},
\end{equation}
where the surface term has been neglected and where we do not consider magnetic fields.
%
This leads to the following collapse criterion
\begin{eqnarray}
\label{eq:virial_collapse}
-E_{pot}> 3 M c_\mathrm{s}^2  + M \sigma^2 ,
\end{eqnarray}

Without turbulence ($\sigma\approx 0$), we are in the regime of Jeans collapse and we recover the definition of the thermal Jeans mass and length. 
Throughout this work, we will use the following generalized definitions for the free-fall time $\tau_\mathrm{ff}$, the Jeans length $\lambda_\mathrm{J}$ and the Jeans mass $M_\mathrm{J}$ to normalise the various expression:
\begin{eqnarray}
\label{eq_tonorm}
    M_\mathrm{J} &=& a_\mathrm{J1} \frac{c_s^3}{G^{3/2} \rho^{1/2}},\\
    \lambda_\mathrm{J} &=& a_\mathrm{J2} \frac{c_\mathrm{s}}{\sqrt{G \rho}}, \\
    \tau_\mathrm{ff} &=& a_\mathrm{J3} \frac{1}{\sqrt{G \rho}},
\end{eqnarray}
where the factors $a_\mathrm{J}$ incorporate numerical constants. 
When needed we will assume that $a_\mathrm{J1}= \pi^{5/2} / 6$, $a_\mathrm{J2}=\pi^{1/2}/2$
and $a_\mathrm{J3}=(3 \pi / 32)^{1/2}$. With this definition the 
Jeans length is half its usual value. The reason of this choice is that our variable $R$ is the radius of the clouds whereas the Jeans mass is usually computed assuming that the cloud diameter is equal to $\lambda_\mathrm{J}$.



Another useful quantity is the virial parameter $\alpha_\mathrm{vir}$. In the case of a uniform spherical cloud it can be defined as
\begin{equation}
    \alpha_\mathrm{vir} = \frac{2 E_\mathrm{kin}}{|E_\mathrm{pot}|}
    = \frac{5 \sigma^2}{4G R^2 \rho_0}.
\end{equation}

\subsubsection{Definition of the star formation rate}

The SFR describes how much gas is converted into stars per unit time. 
\begin{equation}
\nonumber
\mathrm{SFR} = \frac{\mathrm{d} M_*}{\mathrm{d} t}.
\end{equation}
One can define a dimensionless SFR, as introduced by \citet{krum2005}, which quantifies the fraction of the cloud mass converted into stars during one cloud free-fall time:
\begin{equation}
    \mathrm{SFR}_\mathrm{ff} = \dfrac{\mathrm{SFR}}{M_0} \, \tau_\mathrm{ff,0}.
\end{equation}


\subsection{Star formation above a critical density}
\label{PN}

As an approximation, we can break up the definition of the SFR into two parts: the size of the gas reservoir from which stars form $M_\mathrm{reservoir}$, and the typical time-scale that controls the transformation from gas to stars $\tau_\mathrm{cont}$.
The SFR is then simply
\begin{equation}
\mathrm{SFR}_\mathrm{ff} \approx \frac{M_\mathrm{reservoir}}{M_0} \frac{\tau_\mathrm{ff,0}}{\tau_\mathrm{cont}}.
\end{equation}

To estimate the mass reservoir, the \citet{krum2005} (henceforth KM) and also \citet{padoan2011} (henceforth PN), assume that star formation occurs above a critical density $\rho_\mathrm{crit}$.
The mass fraction of gas which has a density above this critical density can be determined by integrating the density PDF weighted by $\rho/\rho_0$:
\begin{equation}
\label{eq:mass_from_pdf}
    \frac{M_\mathrm{reservoir}}{M_0} = \int^\infty_{\delta_\mathrm{crit}} \, \frac{\rho}{\rho_0} \, {\cal P}({\delta}) \mathrm{d}\delta.
\end{equation}
Furthermore, they assume an additional constant efficiency $\epsilon$ with which the gas of collapsing prestellar cores is converted into stars. 
This is motivated by theoretical and observation work that suggest $\epsilon \simeq 0.3-0.5$ \citep[e.g.][]{matzner2000,ciardi2010, andre2010}.

An estimate for the critical density has to be made.
The theory of KM assumes that turbulent support will prevent star formation on scales larger than the sonic length, and so $\rho_\mathrm{crit}$ can be determined from $\lambda_\mathrm{J}(\rho_\mathrm{crit}) \approx \lambda_\mathrm{s}$.
This leads to 
$\rho_\mathrm{crit, KM} = \phi_x \, \rho_0 (\lambda_\mathrm{J0}/\lambda_\mathrm{s})^2$
where $\phi_x$ is a fudge factor of order unity and $\lambda_\mathrm{J,0}$ is the Jeans length at the mean cloud density.
Assuming a Larson relation (Eq.~\ref{eq_turb}) with $\eta_v=0.5$ the sonic length can be expressed as:
\begin{eqnarray}
\lambda_\mathrm{s} = R_0 \left(\dfrac{c_\mathrm{s}}{\sigma_0}\right)^2 = R_0 {\mathcal M}^{-2}.
\end{eqnarray}
The critical density is then
\begin{eqnarray}
\rho_\mathrm{crit, KM} = \pi \frac{ c_\mathrm{s}^2 {\mathcal M}^4}{G \, R_0^2} = \frac{4}{5}\pi \rho_0 \alpha_\mathrm{vir} \mathcal{M}^2.
\label{rho_crit}
\end{eqnarray}
%
%
Thus, the density threshold above which stars are assumed to form scales as $\propto {\mathcal M}^4 $, implying that only very dense and small scale gas contributes to star formation in very turbulent regions.

The critical density is estimated somewhat differently in the PN model. They consider that compressible turbulence creates a network of shocks. $\rho_\mathrm{crit}$ is then obtained by setting the Bonnor-Ebert mass  equal to the mass contained in the typical thickness of the shocked layer. The later can be found by combining the isothermal shock jump conditions and the velocity scaling from Eq.~(\ref{eq_turb}). This results in
\begin{equation}
    {\rho}_\mathrm{crit, PN} \simeq 0.067\, \theta^{-2} \rho_0 \alpha_\mathrm{vir} {\cal M}^2,
\end{equation}
where $\theta\approx 0.35$ is the ratio of the cloud size over the turbulent integral scale (Eqs.~(8-9) of PN).
Thus apart from a numerical factor, the choice of PN and KM regarding $\rho_\mathrm{crit}$ are 
identical. 

The control time scale over which the conversion from gas to stars takes place is chosen to be the time needed for a self-gravitating fluctuation to be replenished, which KM estimated to be a few times the free-fall time at the critical density: $\tau_\mathrm{cont} = \phi_\mathrm{t} \tau_\mathrm{ff,crit}$.
This leads to the following general expression for the SFR (equivalent to Eq.~(19) of KM):
\begin{eqnarray}
\mathrm{SFR}_\mathrm{ff} &=& \frac{\epsilon}{\phi_t } \dfrac{\tau_\mathrm{ff,0}}{\tau_\mathrm{ff,crit}}
\int ^\infty _{\delta_\mathrm{crit}}
\, \frac{\rho}{\rho_0} \, {\cal P}({\delta}) \mathrm{d} \delta.
\label{eq:sfr_simple}
\end{eqnarray}
KM further assume that $\tau_\mathrm{ff,crit} \simeq \tau_\mathrm{ff,0}$, the later being the freefall time on the scale of the entire cloud.
PN, on the other hand, use $\tau_\mathrm{ff,0} / \tau_\mathrm{ff,crit} = \sqrt{\rho_\mathrm{crit}/\rho_0}$.

Using the log-normal density PDF, their respective estimated for the critical density and the relevant time scale, the expressions for the SFR in the formulation of KM and PN become

\begin{eqnarray}
\label{eq:SFR_KM}
\mathrm{SFR}_\mathrm{ff,KM} &=& \dfrac{\epsilon}{2 \phi_t} 
\left[1 + \text{erf}\left( 
\dfrac{S_\delta -2 \ln ({\rho}_\mathrm{crit}/\rho_0 )}{\sqrt{8 S_\delta}}\right)\right], \\
\label{eq:SFR_PN}
\mathrm{SFR}_\mathrm{ff,PN} &=& \dfrac{\epsilon}{2 \phi_t} 
\left[1 + \text{erf}\left( 
\dfrac{S_\delta -2 \ln ({\rho}_\mathrm{crit}/\rho_0 )}{\sqrt{8 S_\delta}}\right)\right]\sqrt{\frac{\rho_\mathrm{crit}}{\rho_0}}.
\end{eqnarray}
In the following and for the sake of simplicity, $\epsilon / (2 \phi _t)$
is assumed to be equal to 1. 

\subsection{Multi-freefall formalism}
\label{sfr_simp}

It is argued that the SFR is obtained by computing the amount of gravitationally unstable mass divided
by the freefall time. Since the freefall time varies with density, using the local 
freefall time rather than a global one seems logical. 
This can be taken into account by moving the freefall time into the integral, as was suggested by \citet{hc2011} (henceforth HC11):
\begin{eqnarray}
\label{sfr_corrected}
\mathrm{SFR}_\mathrm{ff}
 &=& \frac{\epsilon}{\phi_t}
\int ^\infty _{\delta_\mathrm
{crit}} \frac{\tau_\mathrm{ff,0}}{\tau_\mathrm{ff}(\rho)} \frac{\rho}{\rho_0} 
\mathcal{P}(\delta) d{\delta}.
\end{eqnarray}
With the log-normal density PDF, this becomes
\begin{eqnarray}
\mathrm{SFR}_\mathrm{ff} &=& \dfrac{\epsilon}{2 \phi_t} 
\left[1 + \text{erf}\left( 
\dfrac{S_\delta - \ln ({\rho}_\mathrm{crit}/\rho_0 )}{\sqrt{2 S_\delta}}\right)\right] \, \exp{ \left( \frac{3}{8}S_\delta \right)}.
\end{eqnarray}
This SFR is larger than the ones given by Eq.~(\ref{eq:sfr_simple}),
as discussed in \citet{hc2011} and \citet{fk2012}. 

\subsection{Star formation from gravitationally unstable clumps}

In the gravo-turbulent picture, the origin of the self-gravitating structures are density fluctuations generated by turbulence which were large enough for gravity to take over and triggering collapse which sequentially isolates the structure from the more diffuse background.
Let the mass spectrum of collapsing objects be denoted as $\mathcal{N}(M)$, that is the number of self-gravitating clumps with a mass between $M$ and $M+\mathrm{d}M$ is equal to $\mathcal{N}(M) \, \mathrm{d}M$.
The SFR can then be calculated from 
\begin{equation}
\mathrm{SFR}_\mathrm{ff} = \frac{\epsilon}{\phi_t} \int_0^{M_\mathrm{sup}} \dfrac{\tau_\mathrm{ff,0}}{\tau_\mathrm{cont}(M)} \frac{M}{M_0} \mathcal{N}(M) \,dM.
\label{sfr}
\end{equation}
This is the general idea behind the SFR derivation from HC11.
In the spirit of the multi-freefall formalism, each structure mass has its own control time. 
This expression is similar to Eq.~(\ref{sfr_corrected}), except that the bounds of the integral do not need to be chosen. The mass of the system is finite, which naturally limits the mass of unstable structures. In fact, we do not expect turbulence to be able to generate structures with sizes larger than a certain fraction $y_\mathrm{cut}$ of the turbulence driving scale $L_\mathrm{i}$. $M_\mathrm{sup}$ is then the mass corresponding to this maximum structure size $y_\mathrm{cut} L_\mathrm{i}$. For now we will simply assume $y_\mathrm{cut}=1$, but we will come back to this later.

This approach a priori avoids the need for a somewhat artificial critical density.
The complexity now lies in determining the mass spectrum of collapsing overdensities $\mathcal{N}(M)$.

\subsubsection{The Hennebelle \& Chabrier mass spectrum}

An expression for the mass spectrum of collapsing structures is derived in \citet{HC08} (henceforth HC08) using the Press-Schechter formalism developed for cosmology \citep{PS74}. Here, we summarise the key principles. For more details, we refer to HC08.

We assume a density field which is smoothed over a scale $R$.
As before, it is considered that the gas with density higher than a certain critical density $\rho_\mathrm{crit}(R)$ will be gravitationally unstable.
This threshold results from the virial theorem in Eq.~(\ref{eq:virial_collapse}) and is now scale-dependent. 
Equivalent to Eq.~(\ref{eq:mass_from_pdf}), the total mass which is gravitationally unstable at a scale $R$ is
\begin{eqnarray}
\label{gauche}
M_{\rm tot}(R)  = M_0 \int ^{\infty} _ {\delta_\mathrm{crit}(R)} e^{\delta} {\cal P}_R(\delta)  d\delta.
\end{eqnarray}
where we substituted $\rho = \rho_0 e^\delta$ which follows from the definition of $\delta$, and the density PDF is now also scale-dependent.
We can understand intuitively that the density contrast will become less pronounced when smoothing the field, resulting in a narrower PDF.

The mass reservoir in Eq.~(\ref{gauche}) consists of individual collapsing structures of size $R$, each with an average density above the critical density. The mass in each structure is larger or equal to the corresponding critical mass $M_\mathrm{crit}(R) = a_g \rho_\mathrm{crit} R^3$.
On scales smaller than $R$, we would find a number of substructures of higher densities and smaller masses.
Due to mass conservation, the sum of substructures is equal to the total structure mass.
This leads to 
\begin{eqnarray}
M_{\rm tot}(R)  = \int _0 ^ {M_\mathrm{crit}(R)} \, M {\cal N} (M)\,dM. 
\label{droit}
\end{eqnarray}

By equating Eq.~(\ref{gauche}) and~(\ref{droit}), and taking the derivative with respect to $R$, we obtain the general expression
\begin{align}
\label{n_general}
&\mathcal{N}(M_\mathrm{crit}) = \mathcal{N} _1 + \mathcal{N} _2 = \\ 
&\dfrac{\rho_0}{M_\mathrm{crit}} 
\dfrac{dR}{dM_\mathrm{crit}} \, \times  
\left( -\dfrac{d \delta_\mathrm{crit}}{dR} \exp(\delta_\mathrm{crit}) \mathcal{P}_R(\delta_\mathrm{crit}) + 
\int _  {\delta_\mathrm{crit}}^\infty \exp(\delta) \dfrac{d \mathcal{P}_R}{dR} d\delta \right),
\nonumber 
\end{align}
which is Eq.~(33) in HC08.

The expression presents two contributions, respectively $\mathcal{N} _1$ and  $\mathcal{N} _2$.
$\mathcal{N} _1$ is
positive since the critical density decreases for larger scales which reverts back the minus sign. This term would entirely provide the mass spectrum of self-gravitating clumps if the 
density PDF had no scale dependence. 
The second term, $\mathcal{N} _2$,  is usually negative, that is reducing the amount of clumps, and 
precisely comes from the density PDF dependence on scale.
Indeed, the density PDF depends on $S_\delta$, the variance of $\ln \rho$, which is scale dependent. This dependence plays a major role in the classical analysis by \citet{PS74}. To calculate it, it is necessary 
to define a window function in the Fourier space, $W_k$. This leads to
\begin{eqnarray}
\nonumber
S_\delta(R) &=& \int ^\infty _{2 \pi /L_i} \widetilde{\delta}^2(k) W_k^2(R)\, d^3k \\
 &=& S_\delta(L_i) \left( 1 - \left( \dfrac{R}{L_i} \right)^{n_d-3} \right),
\label{eq:S_R}
\end{eqnarray}
where $S_\delta(L_i)$ is typically given by Eq.~(\ref{eq:S_from_mach}) and where 
the window function has been assumed to be sharply truncated in the Fourier space.

In their appendix B, HC08 calculate this term for the log-normal PDF and conclude it only becomes significant for structures whose size is similar to the size of the system.
This term thus implements a cut-off for large masses.
It was neglected in the further steps of the computation in HC08 as well as in HC11 where they use the mass spectrum obtained in HC08 to derive the SFR. Instead, they chose to simply truncate the mass spectrum at an arbitrary fraction of the total cloud mass, setting $y_\mathrm{cut} \approx 0.1$.

\subsubsection{Collapse criterion and normalisation}
The critical mass in obtained from the virial theorem. Combining Eq.~(\ref{eq:virial_collapse}) with the Larson relation Eq.~(\ref{eq_turb}), results in
\begin{equation}
\label{eq:mass_crit_R}
    M_\mathrm{crit}(R) = \frac{ a_v}{G}  R \left(3 c_s^2 + \sigma_\mathrm{0} \left( \frac{R}{R_\mathrm{0}} \right) ^{2\eta_v} \right).
\end{equation}
This can be inverted to obtain $R(M_\mathrm{crit})$.
The critical density is then
\begin{equation}
\label{eq:rho_crit_R}
    \delta_\mathrm{crit} = \ln{ \left( \frac{\rho_\mathrm{crit}}{\rho_0} \right)} 
    = \ln{ \left( \frac{M_\mathrm{crit}}{\rho_0 a_g R^3}\right)}.
\end{equation}
Remark that through the use of the Larson relation in Eq.(\ref{eq:mass_crit_R}), a dependence on the parameter $\eta_v$ is introduced into the mass spectrum.

In their derivations, HC08 normalise the equations by adopting the quantities
\begin{eqnarray}
\nonumber
    \widetilde{R} &=& R / \lambda_\mathrm{J}({\rho_0}), \\
    \widetilde{M} &=& M / M_\mathrm{J}({\rho_0}),
\end{eqnarray}
that is the size and mass divided by, respectively, the Jeans length and mass at the average density.
We will adopt this normalisation also in this work.
This leads to a simplified formulation of the critical mass and density:
\begin{eqnarray}
\label{crit_Mtot_norm}
\widetilde{M}_\mathrm{crit} &=&
 \widetilde{R} +  \mathcal{M}_* ^2 \widetilde{R} ^{2\eta_v + 1}, \\
\delta_\mathrm{crit} &=& \ln \left( \dfrac{\widetilde{M}}{\widetilde{R}^3} \right),
\label{rho_norm}
\end{eqnarray}
where $\mathcal{M}_*$ is 1D Mach number measured at the Jeans length for the average cloud density:
\begin{equation}
\label{eq:Mstar}
    \mathcal{M}_* = \dfrac{\sigma(\lambda_J(\rho_0))}{\sqrt{3} c_s}.
\end{equation}

The resulting HC08 mass spectrum will be discussed in Sect.~\ref{fig:mass_spectrum} as further discussions are needed 
regarding the density PDF.

\subsection{Comparison between theories and simulations}

A systematic investigation between the various theories and a set of simulations that 
include self-gravity and sink particles have been performed by \citet{fk2012}.
About 34 simulations are performed with different Mach numbers, ranging from 3 to 50, and 
compressibilities that go from purely solenoidal to purely compressible.
The approach was to consider model parameters such as $\phi _t$ 
or $y_\mathrm{cut}$ as fudge factors to adjust the analytical values of the SFR to the numerical ones. 
It has been found that once these parameters have been adjusted, a reasonably good agreement
(with $\chi^2 \simeq 1.3$) 
could be obtained for KM and PN theories when the multi-freefall correction is employed.
A relatively less good match (with $\chi^2 \simeq 5$)  was obtained for HC\footnote{Besides, their fit was using values of $y_\mathrm{cut} > 1$ which does not make sense in the HC model framework.}.

\subsection{Revisions to be made}

The SFR estimates presented above make a series of assumptions which can be improved upon. As discussed above, the behaviour observed 
at high Mach numbers in the simulations presented in 
\citet{brucy2020} and \citet{brucy2023} as well as in \citetalias{brucyInefficientStarFormation2024} are not in good agreement with the existing models. 
We identified three crucial points where revisions are needed to construct a more realistic SFR model. Each of these points will be treated in detail in the next sections.

The first is to use a more realistic shape for the density PDF.
The comparisons between the log-normal PDF and the PDF obtained from numerical simulations has revealed good agreement for moderate Mach numbers \citep{kritsuk2007}. 
However, for large Mach numbers the agreement is much 
less satisfactory, particularly regarding the amount of dense gas \citep{federrathComparingStatisticsInterstellar2010,hopkinsModelNonlognormalDensity2013,squireDistributionDensitySupersonic2017,moczMarkovModelNonlognormal2019}. 
This constitutes a problem for SFR theories, because the amount of dense gas plays a central role.
In this work, we will use the functional form provided by \citet{hopkins_pdf_2013}, who transposed the PDF obtained by \citet{castaing1996} to describe the velocity PDF of intermittent incompressible flows. We name this updated PDF the Castaing-Hopkins PDF.
Confronting their formula with a large suite of numerical simulations, 
\citet{hopkins_pdf_2013} found that it provides excellent and robust fits, 
including for the dense gas. 
An attempt to adapt a more realistic density PDF in analytical SFR models was previously made by \cite{moczMarkovModelNonlognormal2019}. However, they used a very simple single-freefall density threshold based model, which lead to very different  conclusions than the ones presented here. 

The second revision concerns the characteristic time over which stars form.
The time it takes for a parcel of dense gas to collapse onto a star is roughly the freefall time.
Most models take into account the fact that the mass reservoir needs to be replenished before the same amount of gas can collapse again during the next freefall time.
However, the estimate for this replenishment time is somewhat arbitrarily set to a few times the freefall time.
In this work, we provide a more accurate estimate for the replenishment time based on the density PDF, which is assumed to be generated by turbulence.
This estimate is then compared to the free-fall time to determine which timescale is limiting star formation. It is argued that whereas the dense gas is associated to small spatial scales, it nevertheless takes a large scale crossing time to rejuvenate it. 

The third  improvement is to take into account the spatial distribution of the gas reservoir. 
In the existing SFR theories, the only link to the density field is through the density PDF,
a one-point statistical quantity.
This implies that these theories would predict the same SFR if the dense gas were organized in a single 
massive clump or, on the contrary, were fragmented into many small mass clumps, as long as the density PDF remains globally the same. 
Intuitively, we expect these cases to experience different behaviour. For instance, if the system contains only small clumps which have a mass below their local Jeans mass, the SFR should be zero. 
We note that the general expression for the mass spectrum of self-gravitating clumps derived by HC08 does depend on the spatial correlation of the density field.
Indeed, the density variance $S_\delta$ which sets the width of the density PDF and is linked to the power-spectrum of $\ln \rho$ is scale dependent.
This comes into play in the second term in Eq.~(\ref{n_general}) which contains the spatial derivative of a scale-dependent density PDF.
This term has been neglected in the SFR derivation of HC11. 
In this work, the starting point of our derivation will be Eq.~(\ref{sfr}) and Eq.~(\ref{n_general}) including the spatial dependence 
of the $\ln \rho$ variance, $S_\delta$. We pay attention to the indice 
of $n_d$ of the power-spectrum of $\ln \rho$.

These three elements are developed in the three following sections. We then discuss the results of the full new model in Sections~\ref{SFR_int} and~\ref{sec:model_params}.

\section{An improved density PDF}
\label{sec:pdf}

\begin{figure}
\includegraphics[width=8cm]{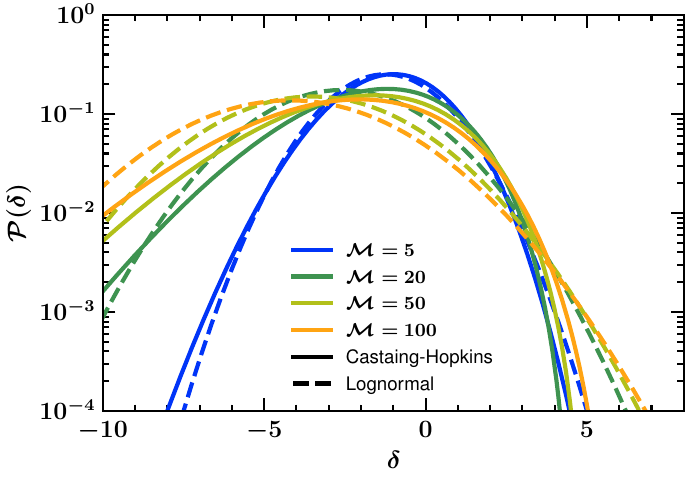}
\caption{The density PDF for three Mach numbers. Solid lines portray the 
Castaing-Hopkins distribution as stated by Eq.~(\ref{eq:hopkins_pdf}), and dashed lines correspond to the 
log-normal PDF as stated by Eq.~\ref{Pr0}. }
\label{pdf_rho}
\end{figure}

Here we describe another density PDF that has been proposed and found to be in good agreement with simulation results. We 
also compare it with the log-normal PDF and discuss the impact it has on some of the existing SFR theories. 

\subsection{Formulation}

It was noted that the density PDF often does not have a perfectly log-normal shape, particularly at high Mach \citep[see e.g.][]{fk2012}.
On the other hand as recalled above, \citet{hopkins_pdf_2013} showed that the PDF obtained by \citet{castaing1996} provides very good fit 
for the published density PDF \citep[see also][]{fedban2015}. This result is confirmed in our companion paper.
We define
\begin{align}
    \label{eq:u_and_lambda}
    u &= \dfrac{\lambda}{1 + T_\mathrm{CH}} - \dfrac{\ln \rho}{T_\mathrm{CH}}, \\
    \lambda &= \dfrac{S_\delta}{2 T_\mathrm{CH}^2}.
\end{align}
The proposed  improved functional form for the density PDF is given by
\begin{equation}
\label{eq:hopkins_pdf}
    {\cal P}(u) = \begin{cases}
    \dfrac{1}{T_\mathrm{CH}}  \displaystyle\sum_{m=1}^{\infty}
    \dfrac{\lambda^m e^{-\lambda}}{m!} 
    \dfrac{u^{m-1} e^{-u}}{(m-1)!} & \text{ if }  u \geq 0,\\
    0 & \text{ if }  u < 0.
 \end{cases}
\end{equation}
In this expression, $S_\delta$ is the variance of $\ln \rho$
and $T_\mathrm{CH}$ is a 
free parameter that described how much the function resembles a log-normal, which is recovered in the limit $T_\mathrm{CH} \rightarrow 0$. 
$T_\mathrm{CH}$ has been envisioned by Castaing as being 
the {\it temperature of the turbulence} by analogy with 
thermodynamics. 
Its constancy through scales represents one of the major assumptions that led to the functional stated by Eq.~\eqref{eq:hopkins_pdf}.

The value of the PDF is entirely determined by the parameters $S_\delta$ and $T_\mathrm{CH}$. For the purpose of this paper, we use Eq.~\eqref{eq:S_from_mach} to estimate the value of $S_\delta$ from the Mach number $\mathcal{M}$ and we use the following formula for $T_\mathrm{CH}$:
\begin{equation}
\label{eq:T_fit}
    T_\mathrm{CH} = \min(\max( 0.6 \log_{10}(\mathcal{M} / 4, 0.1), 0.5) 
\end{equation}
This formula is a satisfactory approximation of the fitted value of $T_\mathrm{CH}$ in turbulent box simulations as shown in \citetalias{brucyInefficientStarFormation2024}.

A convenient form for Eq.~(\ref{eq:hopkins_pdf}) is given by
\begin{eqnarray}
\label{eq:hopkins_pdf2}
  {\cal P}(u) =  \dfrac{1} {T_\mathrm{CH}}  
  {\sqrt{\dfrac{\lambda}{u}} e^{- \lambda - u} I_{1}\left(2 \sqrt{\lambda u}\right)} , 
\end{eqnarray}
where $I_1$ is the modified Bessel function of the first kind. This is the 
expression that we use to carry out the calculations of the new model.

\subsection{Comparison to the log-normal PDF}

Figure~\ref{pdf_rho} compares the two density PDF formulations for various Mach numbers. To compute the Castaing-Hopkins PDF, we have used $S_\delta = \ln (1 + b^2 
\mathcal{M}^2) $ and a value for $T_\mathrm{CH}$ as in Eq.~\eqref{eq:T_fit}. 
Generally speaking, the two PDF make similar predictions at low 
densities. However, the shape of the two PDFs is 
quite different at high densities and this is  
more pronounced at higher Mach numbers.
Consequently, the log-normal tends to overpredict the amount of very dense 
gas compared to the Castaing-Hopkins PDF.
This, in turn, can drastically affect the estimated SFR.

\subsection{The existing SFR models with the updated PDF}
\label{prob_lognorm_mff}
One can ask themselves what the existing star formation theories predict when replacing the log-normal PDF by the Castaing-Hopkins PDF.
The answer is: an SFR which is practically zero.
The KM and PN theories make use of a Mach number dependent critical density above which the density PDF has to be integrated. 
In the case of the Castaing-Hopkins PDF, we find from Eq.~(\ref{eq:hopkins_pdf}) that if 
\begin{equation}
\nonumber
    \ln (\rho / \rho_0) >   S_\delta / (2 T_\mathrm{CH} (1+ T_\mathrm{CH}) ),
\end{equation}
then ${\cal P} = 0$ and so is the SFR.
We can rewrite Eq.~(\ref{rho_crit}) as $\rho_\mathrm{crit} / \rho_0 = K_\mathrm{crit} \alpha_\mathrm{vir} {\mathcal M}^2$, where $\alpha_\mathrm{vir}$ is the virial parameter and $K_\mathrm{crit}$
is a dimensionless factor on the order of unity. 
Combining the two expressions 
and using $S_\delta = \ln (1 + b^2 {\cal M}^2)$, we find that the SFR is zero when  
\begin{eqnarray}
\nonumber
    2 T_\mathrm{CH} (1 + T_\mathrm{CH}) \ln( K _\mathrm{crit} \alpha_\mathrm{vir} {\mathcal M}^2) > \ln(1 + b^2 {\mathcal M}^2)).
\end{eqnarray}
Typical values are $K_\mathrm{crit} \alpha_\mathrm{vir} \simeq 1$, $b \simeq 0.5$ whereas
$T_\mathrm{CH}$ varies with the Mach number \citep[see Fig.~3 of][Fig.~6 of \citetalias{brucyInefficientStarFormation2024} and Eq.~\eqref{eq:T_fit}]{hopkins_pdf_2013}. For Mach number around 20, $T_\mathrm{CH}$ will be close to 0.4.
These numbers leads to the approximate condition
\begin{eqnarray}
\nonumber
     \ln ({\mathcal M}^2) > \ln \left(1 + \frac{{\mathcal M}^2}{4}\right) 
    \Rightarrow
    \dfrac{3}{4}\mathcal{M}^2  > 1,
\end{eqnarray}
which is easily satisfied.
This implies that when applying the Castaing-Hopkins PDF, which is much more accurate than the usually employed PDF, the {\it classical}
theories of the SFR lead to SFR $=0$. Whereas this conclusion is 
exacerbated by the mathematical limit reached when $u<0$, this nevertheless
shows that the KM and PN density criteria must be revised and further stresses that simply taking a density threshold is not sufficient to build a consistent star formation theory. The comparison between models and the 
influence of the density PDF is further explored and discussed in
Section~\ref{comparison}.

\section{A new estimate for the replenishment time} 
\label{sec:time}
We now discuss the time that should be used to estimate the SFR in Eq.~(\ref{sfr}). 
In most theories, a freefall time estimated at some density is used. However, when a 
turbulent density PDF is used, the picture is not consistent. Since turbulence 
is responsible of setting the PDF, the time necessary to rejuvenate the dense gas should also be controlled by turbulence.
In particular, since the dense gas is obtained by contraction of a large volume of diffuse gas, replacing the dense gas likely requires a
large amount of time. 
The question that needs to be addressed is
how much time does it take for a new piece of gas with the same density to be created by turbulent-induced contraction when a piece of fluid with a given density gravitationally collapses.

\subsection{Analytical expression}

We consider a clump of size $R$ and density $\rho$. This clump was formed by the contraction of a larger clump
of size $R' > R$ and density $\rho' < \rho$.
We assume turbulence is the main process that transforms a regular piece of gas into a collapsing piece of gas.
Assuming that turbulence mainly induces one-dimensional contraction, we have 
$R' / R = \rho / \rho'$. 
The typical time to generate the clump at scale $R$ from the clump at scale $R'$ is linked to the crossing time 
at scale $R'$:
\begin{equation}
    \tau_\mathrm{cross}(R') = \dfrac{R'}{\sigma(R')}.
\end{equation}
With Eq.~(\ref{eq_turb}) and with $R_0=L_\mathrm{i}$, where $L_\mathrm{i}$ is the turbulence injection length assumed to be half the 
size of the system,  we obtain
\begin{equation}
    \sigma(R') = \sigma_0 \left(\dfrac{R'}{L_i}\right)^{\eta_v},
\end{equation}
where $\sigma_0$ is the average velocity dispersion of the entire cloud.
This leads to the crossing time
\begin{equation}
\tau_\mathrm{cross}(R') = \frac{L_i^{\eta_v}}{\sigma_0} (R')^{1-\eta_v} = \frac{L_i^{\eta_v}}{\sigma_0} R^{1-\eta_v} \left(\frac{\rho}{\rho'}\right)^{1-\eta_v}.
\end{equation}
The gas of density $\rho'$ at scale $R'$ is transformed into gas of density $\rho$ at scale $R$ on a time $\tau_{R'}$ with an efficiency $\epsilon(R, R')$.
Combining all this, we obtain the time estimate

\begin{equation}
     \tau_{R, R'} =  \dfrac{\tau_\mathrm{cross}(R')}{\epsilon(R, R') }=  \frac{L_i^{\eta_v}}{ \epsilon(R, R')  \sigma_0} R^{1-\eta_v} \left(\frac{\rho}{\rho'}\right)^{1-\eta_v}.
\end{equation}

Now we must sum over all possible scales $R'$ or densities $\rho'$ from which the clump of size $R$ with density $\rho$ could have formed, to get the mean $\tau$. In this summation, it is necessary to weight 
the time $\tau_{R, R'}$ by the gas mass of density $\rho'$ that exists at scale $R'$. 
This number is proportional to the mass-weighted density PDF $\rho' \mathcal{P}(\rho')$. This leads to 

\begin{align}
\tau_\mathrm{rep}(R) &= \dfrac{\int_{-\infty}^{\ln(\rho)} \tau_{R, R'}  \rho' \mathcal{P}(\rho') \,d\ln(\rho')}
{\int_{-\infty}^{\ln(\rho)} \rho' \mathcal{P}(\rho') \,d\ln(\rho')},\\
&=\dfrac{L_i^{\eta_v}}{\sigma_0} R^{1-\eta_v} \rho^{1-\eta_v} \dfrac{\int_{-\infty}^{\ln(\rho)} \epsilon(R, R') ^{-1} (\rho')^{\eta_v} \mathcal{P}(\rho') \,d\ln(\rho')}
{\int_{-\infty}^{\ln(\rho)} \rho' \mathcal{P}(\rho') \,d\ln(\rho')},
\end{align}
which is the mass-weighted time. The denominator ensures that the weight is normalized to 1.

Estimating the efficiency of the conversion of clumps, which at scale $R'$ have a density $\rho'$, into clumps which at scale $R$ have a 
density $\rho$, is not an easy task, and we make several assumptions. First, we assume that it depends on the scale ratio
$\epsilon(R, R') = \epsilon(R/R')$. 
Second, we must have $\epsilon(x_1 x_2) = \epsilon(x_1) \epsilon(x_2)$, that is to say the efficiency 
between scale $R_1$ to $R_3$ is the product of the efficiencies between scales $R_1$ to $R_2$ and $R_2$ to $R_3$. Third, in the limit $x \rightarrow 1$, we 
must have $\epsilon(x) \rightarrow 1$, whereas in the limit $x \rightarrow 0$, $\epsilon(x) \rightarrow 0$. The solution to these constraints is 
$\epsilon(x) = x^\mu$ with $\mu > 0$ which implies
\begin{equation}
\epsilon\left(R, R'\right) = \left(\dfrac{R}{R'}\right)^\mu = \left(\dfrac{\rho'}{\rho}\right)^\mu.
\end{equation}

From numerical experiments, that is to say by confronting the analytical model with the simulation results, 
we found that $\mu$ is typically small and likely below $0.1$.
Therefore, in the following, for the sake of simplicity, we just assume $\mu=0$ and $\epsilon=1$.
This expression leads to
\begin{equation}
\tau_\mathrm{rep}(R) = \dfrac{L_i^{\eta_v}}{\sigma_0} R^{1-\eta_v} \rho^{1-\eta_v} \dfrac{\int_{-\infty}^{\ln(\rho)} (\rho')^{\eta_v} \mathcal{P}(\rho') \,d\ln(\rho')}
{\int_{-\infty}^{\ln(\rho)} \rho' \mathcal{P}(\rho') \,d\ln(\rho')},
\label{trep}
\end{equation}
which using Eqs.~(\ref{eq_tonorm}) can be normalized as
\begin{equation}
\widetilde{\tau}_\mathrm{rep}(R) = \sqrt{\dfrac{32}{3\pi}} \dfrac{\widetilde{L}_i^{\eta_v}}{{\cal M}} \widetilde{R}^{1-\eta_v} \widetilde{\rho}^{1-\eta_v} 
\dfrac{\int_{-\infty}^{\ln(\widetilde{\rho})} (\widetilde{\rho}')^{\eta_v} \mathcal{P}(\widetilde{\rho}') \,d\ln(\widetilde{\rho}')}
{\int_{-\infty}^{\ln(\widetilde{\rho})} \widetilde{\rho}' \mathcal{P}(\widetilde{\rho}') \,d\ln(\widetilde{\rho}')}.
\label{trep_norm}
\end{equation}

Finally, as already explained, if the gravitational time is longer than the replenishment time, the SFR is controlled by the freefall time,
and therefore the control time is given by $\tau_\mathrm{cont}(R) = \max(\tau_\mathrm{rep}(R), \tau_\mathrm{ff}(R))$.

\setlength{\unitlength}{1cm}
\begin{figure}

\includegraphics[width=\linewidth]{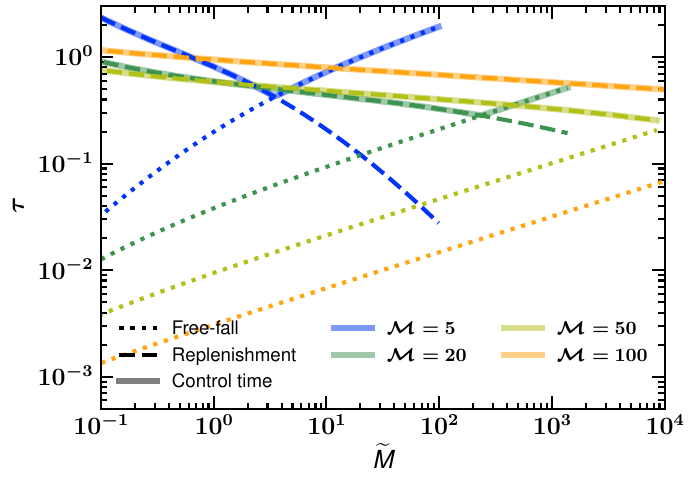}

\caption{Replenishment time $\tau_\mathrm{rep}$  (dashed line), freefall time $\tau_\mathrm{ff}$ (dotted lines) and control time $\tau_\mathrm{cont}$ (solid lines) times for gravitationally unstable structures of mass
$M$ in a system with $b=0.67$, $L_i = 200$,  $n_0$ = 1.5 cm$^{-3}$, and $\eta_d = 0.5$, using the Castaing-Hopkings PDF.  
}
\label{fig:time_mass}
\end{figure}

\subsection{Comparison to the freefall time}

To perform an estimate of the replenishment time, we use a turbulent driving length $L_i$ equal to 200 pc, 
a number density of $n_0$~=~1.5 cm$^{-3}$,
four values of the Mach number and a compressibility described by the $b$ parameter $b=0.67$.
We then use the mass-radius relation, Eq.~(\ref{crit_Mtot_norm}). 
Figure~\ref{fig:time_mass} portrays the replenishment time, the freefall time and the control time as a function of the cloud 
mass. Except at relatively low Mach numbers and for massive clouds, the replenishment time is significantly longer
than the freefall time. The reason is that as explained above, generating high density material is not a local process. It requires 
gathering the gas over large spatial scales. 
For high Mach number ($\mathcal{M} > 20$), the replenishment time is always the controlling time for star formation, and it only weakly depends on the structure's mass.

\section{The mass spectrum of gravitationally unstable clumps}
\label{sec:mass_spectrum}

As discussed in Section~\ref{sec:existing_models}, the HC SFR theory relies on the mass spectrum of collapsing clumps as expressed by the general Eq.~(\ref{n_general}).
In their models, they assume a log-normal density PDF, however
in Section~\ref{sec:pdf} we saw how SFR models may change when using the Castaing-Hopkins PDF instead of the log-normal PDF.
Therefore in this section we investigate how the mass spectrum changes when the Castaing-Hopkins PDF is utilized. 
Another important issue to which we pay attention as well, is the value of $\eta_d$ and $n_d=3 + 2 \eta_d$, the exponent of the power-spectrum of $\ln \rho$, which has usually 
been assumed to be identical to $n_v$ the exponent of the power-spectrum of the velocity.

\subsection{Calculating the mass spectrum}

As discussed above, the mass spectrum, as stated by Eq.~(\ref{n_general}) presents two contributions. 
$\mathcal{N} _1$ is proportional to the density PDF ${\mathcal P}$ and can be easily computed once 
${\mathcal P}$ is known.
To determine the term ${\mathcal N} _2$, one first needs to calculate $d \mathcal{P_R}/dR$ and then perform an integration over densities.
In some rare occasions, the second 
term could become negative, leading to some unphysical features. In practice, 
when it is the case, we simply replace it by 0.

\subsubsection{The mass spectrum for a log-normal PDF}

Assuming a log-normal PDF and using Eqs.~(\ref{n_general}) and (\ref{eq:S_R}), 
 the expression of the 
self-gravitating clump mass spectrum can be inferred. The integral in the term ${\mathcal N} _2$ can be 
explicitly calculated 
 (see Eqs.~(13)-(15) of \citet{HC08}).
 The expression of the corresponding mass spectrum is given in their 
appendix B and is equal to:

{\small 
\begin{align}
\label{big_dens2}
&\widetilde{\mathcal{N}}(\widetilde{M}) \simeq 2\,  
\dfrac{1}{\widetilde{R}^3} \dfrac{1}{1 + (2 \eta_v + 1) \mathcal{M}_*^2 
\widetilde{R}^{2\eta_v}} \\
&\times \left(
\dfrac{1 + (1 - \eta_v)\mathcal{M}_*^2 \widetilde{R}^{2 \eta_v}}{\left(1 + 
\mathcal{M}_*^2 \widetilde{R}^{2 \eta_v}\right)^{3/2}} \right. 
  - \left.
\dfrac{\delta_R^c + \dfrac{S_{\delta}(R)^2}{2}}{\left(1 + \mathcal{M}_*^2 
\widetilde{R}^{2 \eta_v}\right)^{1/2}} \dfrac{n_d - 3}{2}  
\dfrac{ S_{\delta}(L_i)}{S_{\delta}(R)}
\left(\dfrac{\widetilde{R}}{\widetilde{L}_i}\right)^{n_d-3} 
\right) \nonumber
\\ 
&\quad \exp\left(- \dfrac{\left[\ln \left(\dfrac{\widetilde{M}}{\widetilde{R}^3}\right)\right]^2}{S_{\delta}(R)}\right)
\times \dfrac{\exp(-S_{\delta}(R)/8)}{\sqrt{2\pi S_{\delta}(R)}}, \nonumber
\end{align}
}
where $\mathcal{N} = \widetilde{\mathcal{N}}/\mathcal{N}_0$, and $\mathcal{N}_0 = 
\dfrac{\rho_0}{M_0^2}$, $\widetilde{L}_i= L_i / \lambda_{J,0}$, and, unlike 
Eq.~(B1) of \citet{HC08}, we have not assumed that $n_v=n_d$.

\subsubsection{The mass spectrum for a Castaing-Hopkins PDF}

The case of a Castaing-Hopkins PDF is somewhat cumbersome. 
Here we opt for a semi-numerical approach.
The term ${\mathcal N} _1$ in Eq.~(\ref{n_general}) is obtained through the Bessel function formulation
stated by Eq.~(\ref{eq:hopkins_pdf2}). The term ${\mathcal N} _2$ entails $\dfrac{d \mathcal{P}_R}{dR}$.
Using the Bessel function formulation from Eq.~(\ref{eq:hopkins_pdf2}), we get 
\begin{align}
 &{ d {\cal P}_R \over d R} =  \dfrac{\sqrt{\dfrac{\lambda}{u}} e^{- \lambda - u}   }  {T_\mathrm{CH}}
   \left(
 \dfrac{ I_{0}\left(2 \sqrt{\lambda u} \right) + I_{2}\left(2 \sqrt{\lambda u}\right)}{2}   + \right. \\
&\quad \left(  -{ 1 \over T }{d T_\mathrm{CH} \over dR } + \left(  { 1 \over 2 \lambda } -1\right) {d \lambda \over dR } -  \left( {1 \over 2 u  } + 1 \right)  {d u \over dR }  \right)   I_{1}\left(2 \sqrt{\lambda u}    \right)   \Bigg),\nonumber
\end{align}
where we remind that $\lambda= S_\delta /  (2 T_\mathrm{CH}^2 )$ and 
where $I_0$, $I_1$ and $I_2$ are modified Bessel function of the first kind, which can be easily calculated using, for instance, the \textsc{scipy} python library.
As can be noticed, one needs to know $S_\delta$ and $T_\mathrm{CH}$ as well as their 
spatial derivative. We assume that $T_\mathrm{CH}$ varies with $R$ the same way $S_\delta$ does (see Eq.~(\ref{eq:S_R})). 
The values $S_\delta(L_i)$ and $T_\mathrm{CH}(L_i)$ must also be specified and in this paper Eqs.~(\ref{eq:S_from_mach}) and~(\ref{eq:T_fit}) are employed.  
 The final step to get ${\mathcal N}_2$ requires an integration and this is done numerically using the \texttt{quad} function in the library \textsc{scipy}.

An alternative method to estimate ${\mathcal N} _2$ is presented in 
Appendix~\ref{mass_spec_casthop}. It is based on the series formulation as stated by Eq.~(\ref{eq:hopkins_pdf}). 
It does not require further integral calculations but in practice computing the series appears to be time consuming.

\subsection{Description and parameter dependence of the mass spectrum}
\label{mass_spec_ref}

\begin{figure*}
\includegraphics[width=\textwidth]{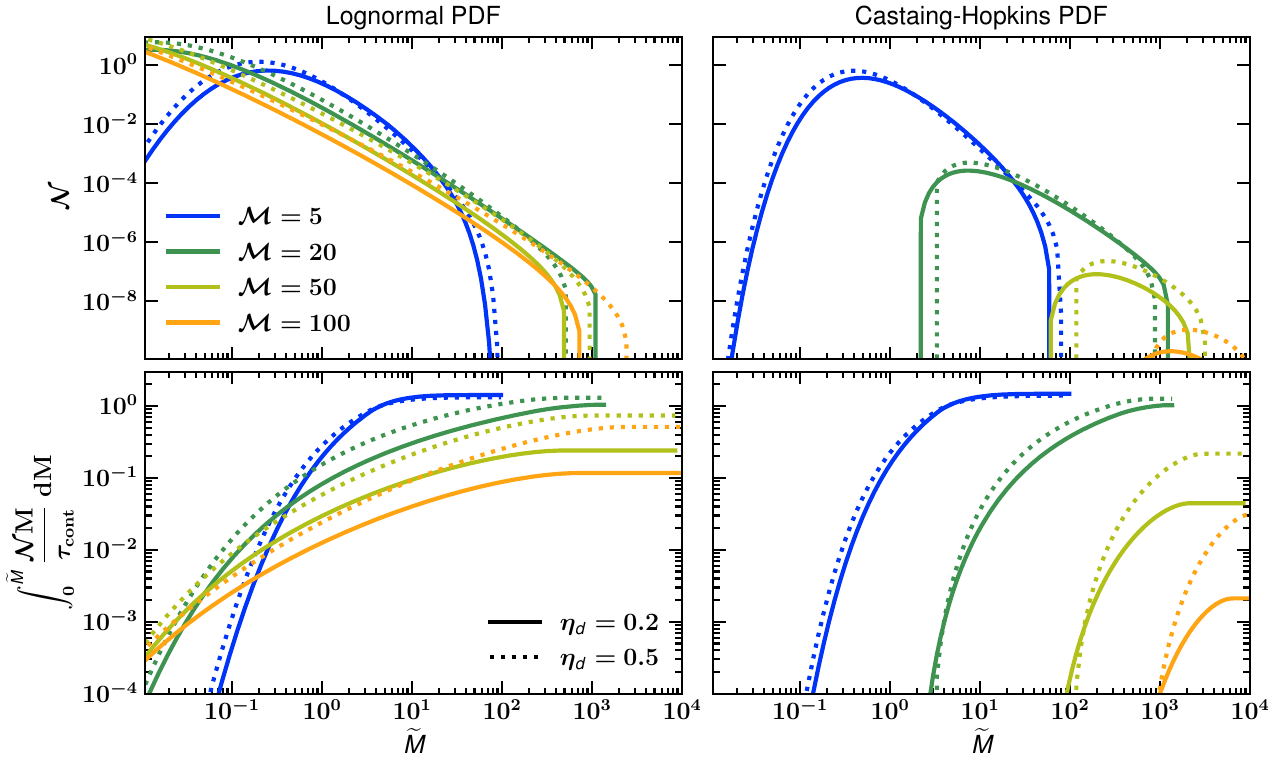}
\caption{
Mass spectra for various Mach numbers and PDF. Top: mass spectrum of self-gravitating structures.
Bottom: time-weigthed cumulative mass spectrum of self-gravitating structures.
Both quantities are plotted for a series of 
Mach numbers from $\mathcal{M} = 5$ (blue) to $\mathcal{M} = 100$ (orange), and for $\eta_d=0.2$ (solid lines) and  $\eta_d=0.5$ (dotted lines).
The models in the left column make use of the log-normal PDF while the ones in the right column use the Castaing-Hopkins PDF. 
For all curves, the compressibility parameter $b$ is 0.67, the injection scale, $L_i$, is assumed to be 200 pc and the mean  density $n_0$ is $1.5$ cm$^{-3}$. 
}
\label{fig:mass_spectrum}
\end{figure*}

\subsubsection{Reference system}
\label{param_model}
Here we study in detail a specific model which has $L_i = 200$ pc, $n_0=1.5$ cm$^{-3}$
and a sound speed  $c_s=0.3$ km s$^{-1}$.
 These values roughly corresponds to the Milky Way which 
present a scale-height of about 100 pc, a mean density around 1 cm$^{-3}$ and a temperature around 10 K for the cold and dense gas.
We remind that what matters is $\widetilde{L}_i = L_i / \lambda_{J,0}$ whose value in the present case 
is about 9. 
We assume that unstable structure can be found up to the the injection scale $L_i$, or, in other words, $y_\mathrm{cut} =1$.
We further assume a value of $b=0.5$ which 
is expected for a mix of solenoidal and compressive modes.
For $\eta_v$ we take 0.4 which would lead to a velocity powerspectrum exponent  of $n_v=3.8$.
For this set of parameters we explore a large range of Mach numbers going from ${\mathcal M} =5$
to ${\mathcal M} =100$. 

Let us recall that the typical velocity dispersion in the Milky-Way is about 6 to 10 km s$^{-1}$ \citep{ferriere2001} leading to Mach numbers, 
for the dense gas of about 20-30. We stress that in the ISM what is usually reported is either the 
Mach number at the scale of the disk scale height of the warm neutral medium or the Mach number at the scale of a cloud for the cold gas \citep[e.g.][]{miville2017}. In both cases, the inferred Mach numbers are smaller than 20 because in the first case the WNM has a sound speed of 8 km s$^{-1}$, whereas in the second case, the cloud scale is smaller than the 
galactic scale height. The quoted Mach numbers of 20-30 are the ones that should be used to describe the
motion of the cold gas at a scale comparable to the disk thickness. 
In high redshift galaxies, much larger velocity dispersion  \citep[e.g.][]{swinbank2012,krumholz2018} have been reported leading to Mach numbers
for the dense gas that can be up to 200.

In the top panel of Fig.~\ref{fig:mass_spectrum} we show the resulting mass spectrum for this reference system when assuming either the log-normal PDF stated by Eq.~(\ref{Pr0}) or the Castaing-Hopkins PDF stated by Eq.~(\ref{eq:hopkins_pdf}).
We also explore two different $\eta_d$ values, 0.2 and 0.5. The choice of these values is motivated by the results from the simulations presented in \citetalias{brucyInefficientStarFormation2024}.

\subsubsection{Dependence on density PDF and Mach number}
The results of the model which uses the Castaing-Hopkins PDF are quite different from what is predicted using the log-normal PDF, especially regarding the behaviour for increasing Mach number.

We see in Fig.~\ref{fig:mass_spectrum} (top) that for ${\mathcal M} =5$, both models predict similar mass spectra, since both PDFs are roughly equivalent at low Mach numbers. 
The mass spectrum presents a maximum around $\widetilde{M} \simeq 1$ and sharply decreases for $\widetilde{M} > 1$. 
For $\mathcal{M} = 20$ and higher, the mass spectrum in the case of the log-normal PDF remains roughly unchanged for increasing $\mathcal{M}$ and follows a power law with an index of $\simeq -1.5$ until it steeply drops around $\widetilde{M} \simeq 10^3$.

However, when a Castaing-Hopkins PDF is employed, an additional cut-off at small masses emerges. This is directly attributable to the fact that the Castaing-Hopkins PDF contains less dense gas compared to the log-normal PDF.
Indeed when the environment is very turbulent, the turbulent dispersion tends to stabilise clouds and fluctuations need to be denser to trigger gravitational collapse. The chance of finding such dense fluctuations is much lower for the Castaing-Hopkins PDF.
The result is that only massive clouds are gravitationally unstable and the typical mass of a collapsing structure increases with ${\mathcal M}$.

\subsubsection{Total amount of collapsing gas}
\label{sec:cumu_mass}

In the bottom panels of Fig.~\ref{fig:mass_spectrum}, we show the cumulative distribution, weighted by the control time, which was discussed in Section~\ref{sec:time}.
Note however that as seen in Figure~\ref{fig:time_mass} this time remains relatively constant for a large range of structure masses.
These figures thus reflect the total mass of gravitationally unstable gas and give information about which structures contain the bulk of this collapsing material.

For low Mach numbers (5 to 20), the amount of self-gravitating gas is roughly independent of Mach number: the curves show a plateau at a normalised value of about 1, both for the log-normal and the Castaing-Hopkins PDF.
For high Mach numbers on the other hand, there is less and less gravitationally unstable gas when the Mach number increases.
The reduction is particularly dramatic when a Castaing-Hopkins PDF is employed as seen in the bottom right panel.
This is a direct consequence of the fact that only massive clouds are gravitationally unstable in the case of the Castaing-Hopkins PDF.
Such clouds are rare, in part due to the limited turbulence injection scale $L_i$ which is assumed to be roughly equal to the size of the astrophysical system. Turbulence cannot generate structures which are larger than the injection scale.

These observations already hint that there are two regimes for star formation depending on the Mach number. From low to intermediate values of ${\mathcal M}$ the total mass of collapsing gas does not depend much on ${\mathcal M}$ and thus the SFR should also not vary strongly with ${\mathcal M}$.
At high Mach numbers however, the unstable structures have a size comparable to the size of the system (or say the turbulent driving scale) resulting in a low amount of collapsing gas and so we can expect the SFR to also drop sharply for increasing ${\mathcal M}$.

\subsubsection{Which clouds contribute to the SFR?}
To understand which clouds contribute most to the reservoir of gravitationally unstable gas, and thus also the SFR, we look at the top and bottom panels together.
On the one hand, we see that the cumulative distributions typically reach their plateau at masses that are somewhat below the extend of the high-mass tail of the mass spectrum. This means that the most massive clouds do not make a large contribution, mostly because such clouds are rare.
We also observe that the contribution of low-mass clouds remains limited to 10\% or less for a log-normal PDF, and is even lower in the case of the Castaing-Hopkins PDF.
The dominant contribution thus comes from clouds of intermediate mass, relatively to the full range of the mass spectrum. The mass of the clouds that contribute the most thus increased with Mach number, especially for the Castaing-Hopkins PDF.

\subsubsection{Dependence on the ln-density power-spectrum}
In Fig.~\ref{fig:mass_spectrum} we also compare the mass spectra obtained for $\eta_d =0.2$ (solid lines) with those for $\eta_d =0.5$ (dotted lines) . We remind that $\eta_d$ is linked to the exponent of the power-spectrum of $\ln \rho$ through Eq.~\eqref{eq:eta_d}. 
Thus the larger $\eta_d$, the more power
there is on large scales implying that the density fluctuations 
tend to be bigger in size than for $\eta_d=0.2$.

For low Mach numbers, the mass spectrum is almost identical for the two values of $\eta_d$.
However for Mach number higher than 20, $\eta_d$ has a significant effect on the shape, particularly at the high mass end. This leads to a strong influence on the total amount of self-gravitating gas, as can be seen in the cumulative distributions in the bottom panels of Fig.~\ref{fig:mass_spectrum}.
For both PDF, there are more massive self-gravitating clouds with $\eta_d=0.5$ than with $\eta_d=0.2$, as we can expect from a steeper power spectrum. 
This is particularly important when a Castaing-Hopkins PDF is employed, since only large clouds are found to be gravitationally unstable. Having additional large scale structures dramatically increases the overall amount of collapsing gas.

\section{Resulting SFR model}
\label{SFR_int}

\begin{figure*}[ht!]
\includegraphics[width=\textwidth]{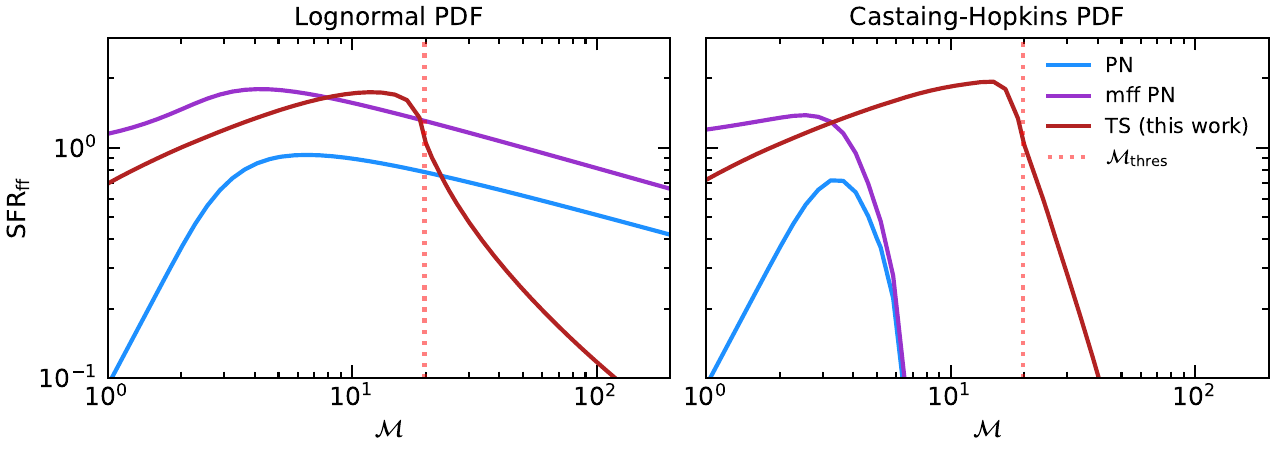}
\caption{
SFR as a function of Mach number. 
The left column is for log-normal PDF and the right column uses the Castaing-Hopkins PDF. 
TS stands for turbulent support (the present model), 
PN and mff PN models are discussed in Sect.~\ref{PN}.
The employed value of $b$ is 0.67, the value of $\eta_d$ is 0.2, the injection scale $L_i$, is assumed to be 200~pc and the mean 
density $n_c=1.5$ cm$^{-3}$. The dotted line is the expected transition between the two regimes according to Eq.~\eqref{eq:criter_Mach}.}
\label{sfr_ref_fig}
\end{figure*}

In the previous sections, we have introduced the more accurate Castaing-Hopkins density PDF, an improved estimate for the replenishment time and have discussed the updated mass spectrum of gravitationally unstable structures. We now study the resulting SFR.

We remind that the SFR is estimated by Eq.~(\ref{sfr}), which sums the contributions of individual collapsing structures weighted by the appropriate characteristic time.
The number of self-gravitating clumps, as generally given by Eq.~(\ref{n_general}), was calculated and discussed in Section 5, both in the case of a log-normal
density PDF (Eq.~(\ref{big_dens2})) and a Castaing-Hopkins PDF (calculated semi-analytically).
The characteristic time, which as explained in section 4, should not be simply the 
freefall time but rather the control time, $\tau_\mathrm{cont}(R)=\max(\tau_\mathrm{rep},\tau_\mathrm{ff})$ where $\tau_\mathrm{rep}$ is the time needed to replace density fluctuations 
with a mass $M_\mathrm{crit}$ at scale $R$. 
The normalised SFR follows
\begin{eqnarray}
\mathrm{SFR}_\mathrm{ff} = \mathrm{SFR} { \tau_\mathrm{ff,0} \over \rho_0 L_0^3 } =  \int _0 ^ {\tilde{M}_\mathrm{sup}} {  \tilde{{\cal N}} ( \tilde{M}) \tilde{M} \over \tilde{\tau} _\mathrm{cont}(R) }d \tilde{M},
\label{sfr_norm}
\end{eqnarray}
where $\tau_\mathrm{ff,0}$ is the freefall time at the mean density $\rho_0$, whereas $\rho_0 L_0^3$ is the total mass of the system.
The bound of the integral, $M_\mathrm{sup}$, is obtained 
through the fact that $R < R_{cut} = y_\mathrm{cut} L_i$ together with 
Eq.~(\ref{eq:mass_crit_R}),
where $y_\mathrm{cut}$ is a dimensionless parameter that is typically below 1. 
The physical meaning of the $y_\mathrm{cut}$ parameter is that in a system 
in which the injection length of turbulence is $L_i$, it seems unlikely to have fluctuations of radius equal to $L_i$. 
The exact value of $y_\mathrm{cut}$ is however not known. When not specified we assume $y_\mathrm{cut}=1$ and we explore its influence in section~\ref{ycut}.

In this section, we investigate the SFR corresponding to the reference system defined in Section~\ref{param_model} with $\eta_d = 0.5$. In the first part, we demonstrate and discuss the emergence of two Mach number regimes. In the second part, we compare the results of our new model to the classical SFR models.
In the next section, we explore the dependence on the system parameters.

\subsection{The two regimes of the SFR}
\label{subsec:tworegimes}

Figure~\ref{sfr_ref_fig} (red line) shows the resulting relation between the SFR and the Mach number for our newly introduced model, which results from 
Eqs.~(\ref{n_general}), (\ref{crit_Mtot_norm}) and (\ref{sfr_norm}).
We name it the turbulent support (TS) model.
As expected from the discussion of the cumulative mass spectrum in Section~\ref{sec:cumu_mass}, we discover the existence of two regimes: for low Mach numbers the SFR is slightly increasing with $\mathcal{M}$, until a critical Mach number above which the SFR drops for increasing $\mathcal{M}$.
The drop is particularly strong in the case of the Castaing-Hopkins PDF, which is due to the strong decrease in the total amount of collapsing gas as seen in the bottom panel of Figure~\ref{fig:mass_spectrum}.

While not the primary cause of the drop, the cut-off term $\mathcal{N}_2$ for the mass spectrum does influence the severity of the SFR reduction at high Mach number. In Appendix~\ref{cut-off} we show that without it, the SFR is a factor of about 3-4 larger in the high Mach regime.

The threshold Mach number which marks the separation between the two regimes is the same for both PDFs.
To obtain a quantitative estimate for the transition Mach number, we can thus use the mathematical expressions for log-normal PDF, which are significantly simpler. 

Mathematically, we see from Eq.~(\ref{big_dens2}) that the mass spectrum is composed
of powerlaw terms and an exponential cut-off, which comes directly from the exponential in the log-normal PDF. An analogous cut-off manifests for the Castaing-Hopkins PDF.
The powerlaw 
part of the mass spectrum corresponds to the density range around the peak of the log-normal PDF. Beyond that point, 
the number of structures drops steeply with increasing size as indeed seen as a high-mass cut-off in the
top panel of  Fig.~\ref{fig:mass_spectrum}. 

To confirm that the high Mach number regime is indeed due to the 
steep decline of the density PDF at high density, we have computed 
the SFR using Eq.~(\ref{big_dens2}) for which the exponential 
term has been set to~1. In this case, it has been found that the SFR 
keeps increasing with the Mach number and that there is no steep decline 
contrary to what it is seen in Fig.~\ref{sfr_ref_fig}.
The exponential term responsible for cutting the cloud powerlaw distribution is
\begin{equation}
\label{eq:exp_term_ln}
E(\widetilde{R}) = \exp \left( - \dfrac{ \left( \ln \left( { \widetilde{M} / \widetilde{R}^3 }  \right) \right)^2}{2 S_\delta(\widetilde{R})} \right)  = \exp\left(-\dfrac{E_N(\widetilde{R})}{2 S_\delta(\widetilde{R})}\right)
\end{equation}
where the scale-dependent variance $S_\delta(\widetilde{R})$ is given by Eq.~\eqref{eq:S_R}.

Using Eqs.~\eqref{eq:mass_crit_R}, \eqref{eq:Mstar} and \eqref{eq_turb}, the numerator $E_N(\widetilde{R})$ becomes
\begin{align}
E_N(\widetilde{R}) &= \left( \ln \left( { \widetilde{M}  /
\widetilde{R}^3 }  \right) \right)^2 \\
&= \left[ \ln \left( {  \widetilde{R}^{-2} \left( 1  + \dfrac{\mathcal{M}^2}{3}\left(\dfrac{\widetilde{R}}{\widetilde{L_i}}\right)^{2\eta_v} \right) }  \right) \right]^2\nonumber
\\
&\approx \left[ \ln \left( { \dfrac{\mathcal{M}^2 \widetilde{L_i}^{2}}{3}\left(\dfrac{\widetilde{R}}{\widetilde{L_i}}\right)^{2(\eta_v -1)} }  \right) \right]^2
\nonumber
\end{align}
when $\mathcal{M}$ and $\widetilde{R}$ are big enough.
The exponential term $E(\widetilde{R})$ will be very small unless $E_N(\widetilde{R}) \ll S_\delta(\widetilde{R})$.
The term $E_N(\widetilde{R})$ is minimal when $\widetilde{R}$ is close to $\widetilde{R}_\mathrm{peak}$\footnote{$\widetilde{R}_\mathrm{peak}$ would have been the maximum of $E(\widetilde{R})$ if $S(\widetilde{R})$ was a constant.} such that 
\begin{equation}
\label{eq:Rmax}
    \widetilde{R}_\mathrm{peak} = \left(\dfrac{\mathcal{M} \widetilde{L_i}}{\sqrt{3}} \right)^{1 - \eta_v} \widetilde{L_i}.
\end{equation}
From Eq.~\eqref{eq:S_R}, we see that the variance $S(\widetilde{R})$ becomes zero in the limit $\widetilde{R} \rightarrow \widetilde{L_i}$. 
As a consequence, the mass spectrum is strongly reduced by the exponential term $E(\widetilde{R})$ at all scales $\widetilde{R}$ unless 
\begin{equation}
\label{eq:crit_Rpeak}
    \widetilde{R}_\mathrm{peak} < \widetilde{L_i}.
\end{equation}
We note that if $y_\mathrm{cut} < 1$, this condition becomes  $\widetilde{R}_\mathrm{peak} < y_\mathrm{cut}\widetilde{L_i}$. Using Eq.~\eqref{eq:Rmax} and assuming $\eta_v < 1$, we find that the condition~\eqref{eq:crit_Rpeak} is equivalent to
\begin{equation}
    \label{eq:criter_Mach}
    \mathcal{M} <  \mathcal{M}_\mathrm{thres}
\text{ where }
    \mathcal{M}_\mathrm{thres} = \sqrt{3}\widetilde{L_i} = \dfrac{L_i \sqrt{3 G \rho_0}}{c_\mathrm{s}}.
\end{equation}
When the Mach number is above $\mathcal{M}_\mathrm{thres}$, the mass spectrum will be considerably attenuated by the term $E(\widetilde{R})$: we enter the high-Mach regime, where star formation start to be inefficient.
For our reference system with $L_i = 200$ pc the above-derived criterion results in $\mathcal{M}_\mathrm{thresh} \approx 20$, in excellent agreement with Fig.~\ref{sfr_ref_fig}.

We can take a step back and look for a physical interpretation of our criterion. Rewriting Eqs~\eqref{eq:criter_Mach}, we see that star formation is efficient only if 
\begin{eqnarray}
{L _i \over \lambda_{J,0} } > \dfrac{\cal M}{\sqrt{3}},    
\label{simple_cond}
\end{eqnarray}
We note that in Eq.~(\ref{simple_cond}), 
 the sound speed $c_s$ appears both in $\lambda_{J,0}$ and in ${\cal M}$. 
By removing it on both sides, we can write that the transition
 between efficient and inefficient star formation occurs when the turbulent Jeans length is comparable to the injection length of turbulence: 
 \begin{eqnarray}
{L _i \over \lambda_\mathrm{J,turb} } \simeq \dfrac{L _i}{{\sigma}/{\sqrt{G \rho_0}} }  \simeq 1,
\label{jeansturb_cond}
\end{eqnarray}
where all numbers of order unity have been dropped. 

This is very interesting because the turbulent Jeans length happens to be the scale where the turbulent velocity dispersion takes the role of the relevant speed to estimate the support against gravity.

Interestingly, the threshold Mach number depends only on the turbulence injection scale and the average density of the system. The details of the turbulence driving mechanism, that is the compressibility as well as the slope of the velocity or density power spectrum that is generated, play no role in setting the transition.

\subsection{Comparison between models}
\label{comparison}

Figure~\ref{sfr_ref_fig} also shows for comparison the PN and multi-freefall PN models (see Sect.~\ref{PN} and \ref{sfr_simp} respectively) as a blue and purple curve.
Clearly the behaviour of these two models is rather different compared to our new TS model (red line). 
At low ${\mathcal M}$, the mff PN and the TS model present comparable values, contrary to the PN model that typically produces lower values.
When a log-normal PDF is employed (left panel) and for $ {\mathcal M} > 3-5$, the SFR predicted by the models PN and multi-freefall PN monotically decrease
with Mach number at a moderate rate, whereas the TS model starts to steeply drop. 

As anticipated in Sect.~\ref{prob_lognorm_mff},  
the difference is even far more dramatic when a Castaing-Hopkins PDF is employed as portrayed in the right panel. 
For $ {\mathcal M} > 6$, due to small amount of dense gas predicted by this PDF, the PN and multi-freefall PN models lead to a vanishing SFR.
Since the Castaing-Hopkins PDF has been found to be in much better agreement with simulation results than the log-normal, 
this constitutes a serious difficulty for this type of model that consists in estimating the SFR from the very dense gas. 

The difference of behaviour between the three models is thus very significant. Let us remind that whereas the PN model consists essentially in estimating 
the very dense gas, gravitationally unstable at the thermal Jeans scale only, the TS model takes into account the gas unstable at all scales, in particular 
the gas supported by turbulence dispersion.

\section{Influence of the model parameters on the SFR}
\label{sec:model_params}
In this section we examine how the various model
parameters such as 
the power spectrum of the 
density field described by $n_d$ or $\eta_d$, 
the maximum structure size described by $y_\mathrm{cut}$, 
the turbulent forcing scale $L_i$ and the mean density $n_0$, affect the relation between SFR and Mach numbers. 
Unless otherwise specified we use the parameters discussed in Sect.~\ref{mass_spec_ref}.

\subsection{Dependence on the compressibility}

\setlength{\unitlength}{1cm}
\begin{figure}
\centering
\includegraphics[width=\linewidth]{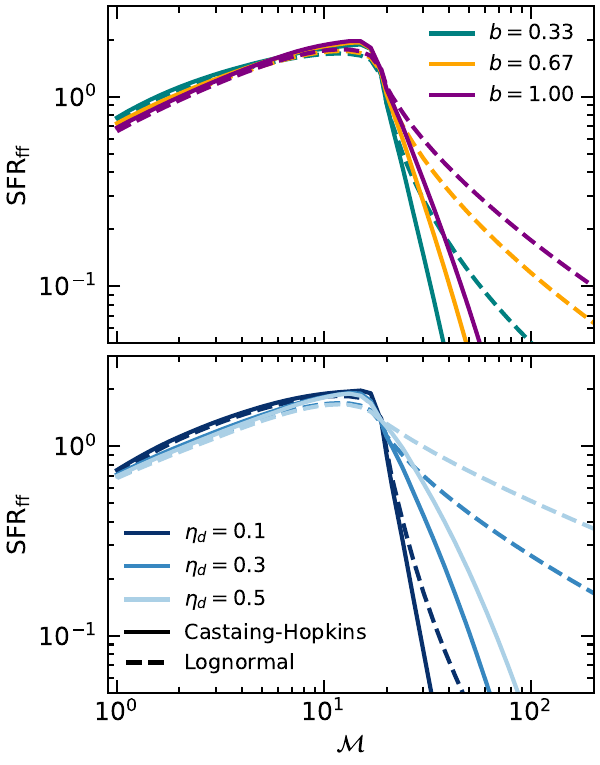}

\caption{Predicted SFR for various parameters. Top: SFR for three values of $b$ as a function
of the Mach number ($L_i = 200, \eta_d = 0.2$).  
Bottom: SFR for three values of $\eta_d$  as a function of the Mach
number ($L_i = 200, b = 0.67$), where we remind that $n_d=3+2 \eta_d$. 
Solid lines represent the model using the log-normal PDF and dot-dashed ones used the 
Castaing-Hopkins PDF.}
\label{sfr_mach}
\end{figure}

Upper panel of Fig.~\ref{sfr_mach} portrays the SFR as a function of Mach number for 
three values of the compressibility parameter $b$ where $S_\delta = \ln (1 + b^2 {\mathcal M}^2)$. The values $b=0.33$, 0.5 and 1 correspond to a purely solenoidal, mixed and purely compressive turbulence 
forcing \citep{Federrath08}. 
The dot-dashed lines have been calculated using a log-normal PDF (Eq.~\ref{Pr0})
and the solid lines used the Castaing-Hopkins PDF (Eq.~\ref{eq:hopkins_pdf}). 

At low Mach number, the influence of $b$ on the SFR remains limited. Only at high Mach number, that is for ${\cal M} > 20$  is there a significant difference. The steep dependence on Mach number in this regime originates from the stabilisation of massive clouds by turbulent support. This implies that 
denser perturbations must occur when the Mach number is larger. The probability of finding a 
denser fluctuation is higher when $b$ is larger.
For the same reason, there is a clear difference between the SFR calculated using the log-normal PDF 
and the Castaing-Hopkins PDF. While for low Mach numbers the Castaing-Hopkins PDF resembles a log-normal, for high Mach numbers the amount of dense gas is significantly reduced resulting in a lower SFR.

\subsection{Dependence on the logarithmic density powerspectrum}

Lower panel of Fig.~\ref{sfr_mach} displays the SFR as a function of Mach numbers for 
three values of $\eta_d$ where we recall that $n_d = 3 + 2 \eta_d$ is the power-law exponent of the power-spectrum of $\ln \rho$. 
While there are almost no changes to the SFR when varying $\eta_d$ in the low Mach regime, there is a very significant dependence for high Mach numbers. The reason is that larger $\eta_d$ correspond
to more numerous large scale density fluctuations which tend to be much more gravitationally unstable 
than smaller scale fluctuations. As discussed previously, the fluctuation distribution is really critical
in the high Mach number regime.

This explains why the value of $\eta_d$ has a strong impact.

In the high Mach regime and for the log-normal distribution, the SFR decreases  significantly more steeply 
when $\eta_d = 0.15$ than for $\eta_d =0.3$ or 0.5. This decrease is considerably steeper for the Castaing-Hopkins
PDF and the dependence on $\eta_d$ is less pronounced than for the log-normal PDF. Again this is a consequence of the 
high density part. The Castaing-Hopkins PDF presents less dense gas than the log-normal and therefore the amount of self-gravitating gas increases
with $\eta_d$ also less rapidly.

\subsection{Dependence on the maximum structure size}
\label{ycut}
Here we explore the influence of the $y_\mathrm{cut}$ parameter, which 
we remind defines the radius of the largest structures, $R < y_\mathrm{cut} L_i$ that are assumed to exist. In practice,
the SFR defined by Eq.~(\ref{sfr_norm}) 
is integrated up to $M_\mathrm{sup}$, where 
$M_\mathrm{sup}$ is the mass associated to $y_\mathrm{cut} L_i$ through Eq.~(\ref{eq:mass_crit_R}).

\begin{figure}
\includegraphics[width=\linewidth]{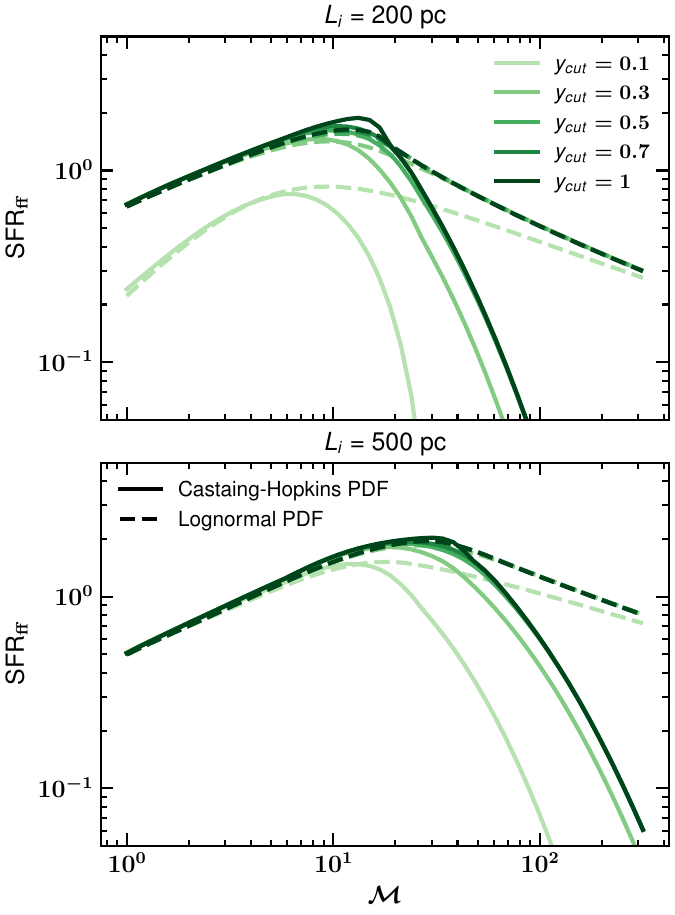}
\caption{SFR for two values of $L_i$, the injection length of turbulence,  and four values of $y_\mathrm{cut}$,  as a function of the Mach
number. 
Solid lines represent the model using the Castaing-Hopkins PDF and dashed ones used the log-normal PDF.
For all curves $\eta_d = 0.5$.
}
\label{sfr_ycut}
\end{figure}

Figure~\ref{sfr_ycut} portrays the SFR as a function of Mach number 
for 4 values of $y_\mathrm{cut}$. For $y_\mathrm{cut} > 0.5$, the SFR does not depend
strongly on it. The main effect of reducing $y_\mathrm{cut}$ is to smooth the transition between the low and high Mach regimes but otherwise the 
value of the SFR remains unchanged. 
This is however different for smaller values. 
For $y_\mathrm{cut}=0.3$, the SFR at large Mach is reduced whereas 
it remains unchanged at low Mach with respect  to the case $y_\mathrm{cut}=1$.
Finally, for $y_\mathrm{cut}=0.1$ and $L_i=200$ pc, the SFR is lower 
at all Mach compared to the case $y_\mathrm{cut}=1$. 

A priori, we do not know what the value of $y_\mathrm{cut}$ should be. We can imagine that the generation of the largest structures by turbulence is a stochastic process, meaning that some system may have a massive structures while other do not. 

\subsection{SFR as a function of mean density and turbulent driving scale}

\begin{figure}

\includegraphics[width=\linewidth]{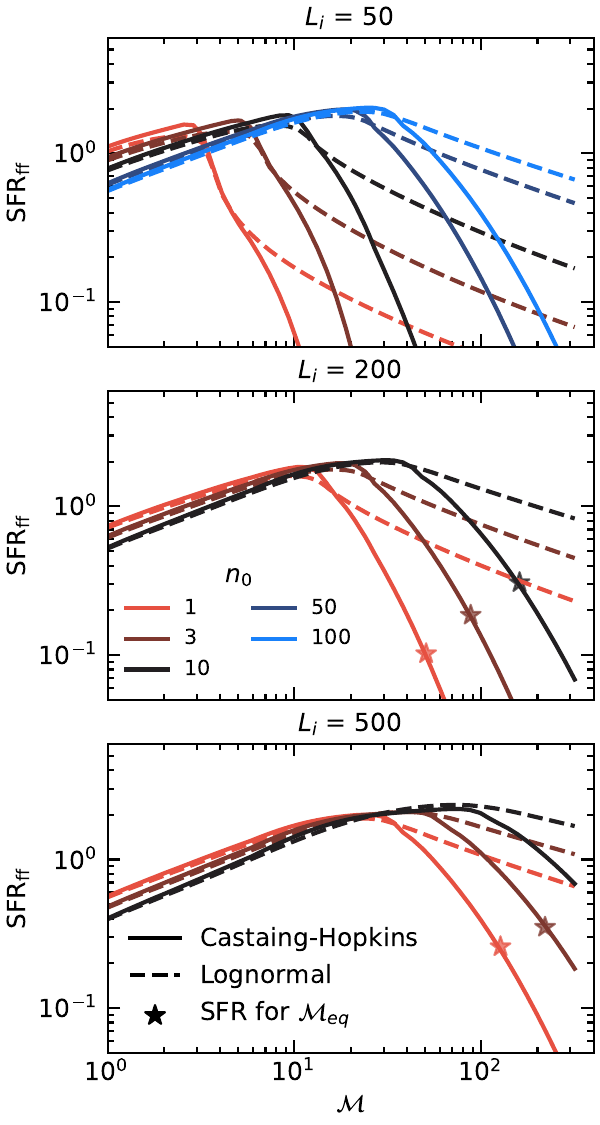}
\caption{SFR for three values of $L_i$, the injection length of turbulence,  and several values of $n_0$, the mean density, as a function of the Mach
number. 
Solid lines represent the model using the Castaing-Hopkins PDF and dot-dashed ones used the log-normal PDF.
For all curves $\eta_d = 0.5$.
The star markers correspond to the Mach number which results from the mechanical equilibrium within a self-gravitating gas rich galactic disk as stated by Eq.~(\ref{equil}). 
}
\label{sfr_n0}
\end{figure}

Here we explore further the dependence of the SFR on the density and driving scale with the goal of providing easy references
to a large number of configurations. 
Figure~\ref{sfr_n0} portrays the SFR as a function of Mach number for 3 values of $L_i$ and several densities. 
We remind that the theory presented above depends on $\widetilde{L}_i = L_i / \lambda_{J,0}$ and that the transition between the low and high Mach regimes
 broadly occurs when  $\widetilde{L}_i \simeq {\cal M} $. 

The top panel displays the case $L_i=50$ pc which broadly corresponds to the size of giant molecular clouds. We focus on  three mean 
densities $n_0=10$, 30 and 100 cm$^{-3}$, which are typical values for molecular clouds. As can be seen the behaviour of the 
SFR is similar to what has been previously inferred with a plateau at low to moderate Mach number followed 
by a steep drop at higher Mach numbers, which corresponds to a very inefficient star formation.  
We see that the  Mach number at which the transition occurs is proportional to $n_0^{1/2}$ as expected from Eq.\eqref{eq:criter_Mach}. 
Interestingly, rather low SFR values are obtained for values of ${\cal M} \simeq $ 20 for $n_0=10$ cm$^{-3}$
to ${\cal M} \simeq $ 50 for $n_0=100$ cm$^{-3}$. These values are typical of what is inferred for molecular clouds. 
For instance assuming that they follow Larson's relations \citep{Larson81,HF12}, we expect a velocity dispersion on the order of 7 km s$^{-1}$ which 
given the assumed sound speed of 0.3 km s$^{-1}$ (relevant for the dense gas), leads to a Mach number of about 20, whereas the mean density at 50 pc is 
about 20 cm$^{-3}$. Consequently, the present theory appears to be entirely compatible with the relatively low 
SFR inferred from observations \citep[e.g.][]{lada2010}. 

The middle panel shows results for  $L_i=200$ pc, which again is comparable with (twice) the Galactic scale height. The densities that we considered
are $n_0=1$, 3 and 10 cm$^{-3}$ is typical to the density averaged at these scale in Milky-Way like galaxies.

Again the Mach numbers that corresponds to low SFR values are comparable 
to the ones observed  in the Galaxy inside the solar circle (velocity dispersion of 10 km$^{-1}$ leading to ${\cal M} \simeq 30$).

The bottom panel portrays the SFR for $L_i=500$ pc and densities equal to  $n_0=1$, 3 and 10 cm$^{-3}$. As expected, the results are similar to the middle panel ones 
once the mach numbers are multiplied by a factor compatible with $500/200=2.5$. The Mach number values are now quite high as getting a low SFR would 
require ${\cal M} \simeq 100$ or even larger depending of the density. Let us remind that very high velocity dispersions have been inferred in early type galaxies where 
velocity dispersions of up to 100~km~s$^{-1}$ have been claimed \citep[e.g.][]{swinbank2012,krumholz2018}.

\section{Discussion}

\subsection{Criteria for inefficient star formation vs galactic vertical equilibrium}

One of the main results of this paper is the existence of a high Mach regime which corresponds to a 
low SFR as explained in Section~\ref{subsec:tworegimes}. 
In essence, if $\widetilde{L}_i < {\cal M} / \sqrt{3} $, most of the
density fluctuations are stabilised by the turbulent velocity dispersion. If galaxies are naturally in this high Mach number regime, it could explain their low observed SFRs.

The condition to be in the inefficient SFR regime presents strong similarities with the conditions for vertical equilibrium 
within a gas dominated galaxy. The gravitational potential, $\phi$, is about 
$\phi \simeq 4 \pi G \rho h^2$, where $h$ is the galaxy scale height. The mechanical equilibrium 
leads to $\phi \simeq \sigma _{1D} ^2$. Using the expression for the Jeans length, we get
\begin{eqnarray}
{\cal M_\mathrm{eq}} =    \frac{\sqrt{3} \pi h}{\lambda_{J,0}}.
\end{eqnarray}
Assuming that $ h \simeq L_i$, that is to say the injection of turbulence is the smallest spatial scale of the system, we see that 
this leads to  
\begin{eqnarray}
\label{equil}
{\cal M_\mathrm{eq}} \simeq   \sqrt{3}   \pi   \widetilde{L} _i  \simeq \pi \mathcal{M}_\mathrm{thres}  , 
\end{eqnarray}
which therefore implies that $\widetilde{L} _i  < {\cal M_\mathrm{eq}}  / \sqrt{3} $ and thus that through vertical mechanical equilibrium, a galactic disk should naturally be in high Mach low SFR 
regime. 
The star markers in Fig.~\ref{sfr_n0} display the equilibrium Mach number $\mathcal{M}_\mathrm{eq}$ corresponding to Eq.~(\ref{equil}) and the expected SFR for a Castaing-Hopkins PDF. They show that a gas rich galactic disk is therefore expected to be in 
high Mach number regime meaning that an SFR$_\mathrm{ff}$ of about 0.3-0.4 could naturally result from the turbulent dispersion. Since in this regime the SFR sensitively 
depends on the Mach number, this estimate remains indicative. A detailed equilibrium and energy injection description is needed to infer 
more quantitative values. In particular, taking into account the stellar contribution for the gravitational potential,
Eq.~\ref{equil} would become 
\begin{eqnarray}
\label{equil2}
{\cal M_\mathrm{eq}} \simeq   \sqrt{3}  \pi    \widetilde{L} _i   \sqrt{1 + { \rho _* \over \rho_g}} , 
\end{eqnarray}
where $\rho _*$ is the mean stellar density, that is the mean density corresponding to the total stellar mass.  
This may indicate that the star markers in Fig.~\ref{sfr_n0} have to be shifted to the right by a factor of about 3 resulting in even lower values of the SFR. 

The conclusion that galactic disks in vertical turbulent equilibrium should naturally present a low star formation rate relies however on a strong assumption: that turbulent injection length is comparable to the disk thickness. 
Density fluctuations arising in the galactic disk plane could be significantly larger than the disk thickness, which would suggest that the relevant value of the turbulent injection scale $L_i$ to consider is also larger than the disk thickness.

\subsection{About considering powerlaw density PDF to calculate the SFR}
It is now well established, both observationally 
\citep{Kainulainen09} and theoretically \citep{Kritsuk11,leeh2018a}
that in regions dominated by self-gravity, the density PDF develops a power-law tail as the result of gravitationnal collapse. 
Typically these power-laws have an exponent of about 3/2. 
This can be understood by the fact that a collapsing envelop 
tends to present a $\rho \propto r^{-2}$ spatial profile.
A uniform in-fall velocity combined with a density profile 
$ \propto r^{-2}$ leads to a constant accretion rate, which is 
probably why this type of solutions tend to appear \citep[see for instance][]{larson1969}.
Let us stress however that if $\rho = A r^{-2}$, then energy conservation 
leads to $v_r = 2 \sqrt{2} \sqrt{\pi G A}$ and thus to an accretion rate 
$\dot{M} \propto A^{3/2}$. However, irrespective of the value of $A$, 
the density PDF is $\propto \rho^{-3/2}$. This implies that the 
power-law tail of the PDF is highly degenerated regarding the SFR and may correspond to very different accretion rates. 
To illustrate this point further, we may for instance consider the 
self-similar isothermal collapse solutions \citep{larson1969,Shu77}. They all 
present densities, $\rho \propto r^{-2}$, which  in term of density PDF  
lead to a power-law tail $\propto \rho^{-3/2}$ but their associated accretion 
rate may vary by orders of magnitude. 
This illustrates well that the power-law tail of the density PDF is 
a consequence of what is happening at larger scales. The larger scales 
are setting the SFR to which controls power-law tail. 
In principle, the latter should therefore be a prediction of the SFR theory. 

Note however, that eventually the choice of considering a
power-law tail to the PDF in order to estimate the SFR 
\citep{burkhart2018} depends of the problem under investigation. 
For instance if ones wants to predict what the SFR within a molecular clouds for which such a power-law tail is observed, it could make 
perfect sense to include it because at the time of interest 
the star forming matter has already been assembled and the evolution time
is not expected to be longer than a few cloud freefall times.

\subsection{Possible further improvements}
There are several possible lines of improvements that need to be explored 
in future studies.
One of them is obviously taking into account the magnetic field which we have ignored here. It was considered in \citet{HC13} and could in principle be added easily. Its effect on the flow statistics (density PDF and power spectrum) should be studied as well (see \citetalias{brucyInefficientStarFormation2024}). 
A recent study by \citet{beattie2022} has revealed that 
there may be significant differences regarding the density PDF between 
hydrodynamics and MHD. This could also affect the value of $\eta_d$ 
for instance. 

Other important improvements would be accounting for stellar feedback, 
which has been found by many groups to significantly affect the SFR. 
This is true at the cloud scales \citep[e.g.][]{Wang10,Nakamura11,Federrath15,verliat2022} in which case it is largely 
due to role of protostellar jets and at the galactic scales 
\citep[e.g.][]{hopkins2011,kim2013,hennebelle2014} where supernovae and HII regions are the dominant processes. 

Similarly, it would be worth to account for temperature variations. This could 
possibly have a significant influence at low Mach numbers albeit the effect is probably limited in the high Mach regime identified in this paper. 

Finally, we emphasize that it would be necessary to better understand and describe 
the transition from the low to the high Mach regimes. As discussed above this transition occurs when the size of the density fluctuations necessary to 
produce a self-gravitating clump is comparable to the size of the system which in turn is assumed to be the injection length of the turbulence. This is somehow 
accounted for by the density variance $S_\delta(R)$ which is presently described 
by Eq.~(\ref{eq:S_R}). This however is most likely too simple to accurately described the distribution of spatial fluctuations comparable 
to the size of the system.

 \section{Conclusion}
 
We have presented a new analytical model to predict the SFR in a turbulent environment. 
Contrary to previous propositions, our approach does not rely on a direct integration of the density PDF over a certain density threshold but rather on an estimate of the density fluctuations which at a given spatial scale are not supported against gravity either by thermal or turbulent support. 
We account for the density spatial distribution which is playing an important role regarding the value of the SFR because it affects the distribution of gravitationally unstable masses.
We also revise the characteristic time that is relevant for the SFR arguing that this is not the freefall time but rather the time to replenish the amount of dense gas. 
Finally we also explore the influence of the density PDF proposed by \citet{castaing1996} and \citet{hopkins_pdf_2013}, which contains significantly less dense gas.
We show that at high Mach number, it leads to substantial differences with the log-normal PDF.

We identify two different regimes of star formation that occur respectively at low and high Mach numbers. 
At low Mach numbers, our model predicts that the SFR is remarkably independent of ${\cal M}$.
However, when the Mach number is higher than a threshold proportional to the injection length of the turbulence divided by the Jeans length, there is a steep decline of the SFR  with ${\cal M}$. 
The reason is that because turbulent dispersion is accounted for in the model, 
the typical gravitationally unstable mass increases with ${\cal M}$. 
In the low Mach regime, increasing the Mach number will lead to the formation of larger and more massive unstable clouds
without significantly changing the total unstable mass in the system.
However, the maximal scale (and, consequently, the maximal mass) at which an unstable structure can be found is limited by the injection scale of the turbulence.
When the Mach number is such that the turbulent Jeans length is larger than the turbulent injection length, only denser density fluctuations may possibly reach the mass needed for gravitational instability. Since according to the density PDF, the number of such fluctuations steeply falls, so does the SFR.

\begin{acknowledgements}
This project was funded by the European Research Council under ERC Synergy Grant ECOGAL (grant 855130), led by Patrick Hennebelle, Ralf Klessen, Sergio Molinari and Leonardo Testi.
The authors acknowledge Interstellar Institute's program "II6" 
and the Paris-Saclay University's Institut Pascal for hosting discussions 
that nourished the development of the ideas behind this work.
\end{acknowledgements}

\section*{Data availability}

The code to compute the TS model and to recreate the plots of this paper is available at \url{https://gitlab.com/turbulent-support/ts_sfr_model}.

\bibliography{lars,global}{}
\bibliographystyle{aa} 
\appendix

\section{SFR for the log-Poisson density PDF: The integral formulation}
\label{mass_spec_casthop}

To obtain the SFR using the Castaing-Hopkins 
 density PDF as stated by Eq.~(\ref{eq:hopkins_pdf}),  it is necessary to 
calculate  the term $d {\cal P}_R / d R$, which appears 
in  Eq.~(\ref{n_general}).

Here we propose an integral formulation, in addition to the Bessel function formulation presented in Section~\ref{sec:pdf},
 which can also be useful although the calculations are much longer than using  
 Bessel functions available from mathematical packages. 
 We have 
 \begin{align}
 \label{dPdR}
    &\dfrac{d {\cal P} _R}{d R}(\delta) = \\
   &\dfrac{1}{T} \sum _{m=1} 
\left( m u - \lambda u + \dfrac{m - 1}{1 + T} \lambda - \dfrac{\lambda u}{1 + T }  \right)     
      \dfrac{d \lambda}{dR} \dfrac{\lambda ^{m-1} e ^{-\lambda}}{m !} \dfrac{u ^{m-2} e^{-u}}{(m-1) !}.
\nonumber
\end{align} 
In Eq.~(\ref{n_general}), the quantity 
\[\int _{\delta_{\mathrm{crit}}} ^ {\infty} d\delta  \exp(\delta) \dfrac{d {\cal P}_R}{d R}\]
appears and needs to be calculated. It itself requires to calculate the integral
\begin{equation}
    I_m = \int _{\delta_{\mathrm{crit}}} ^{\delta_{\mathrm{max}}  } d\delta  \exp(\delta) u^m \exp(-u),
\end{equation}
where  $\delta _{\mathrm{max}}= \dfrac{\lambda T}{1+T}$. Setting $v = (1+T) u$, we obtain 
\begin{eqnarray}
   I _m &=&  \int _{\delta_{\mathrm{crit}}} ^{\delta_{\mathrm{max}}  } d\delta  \exp(\delta) u^m \exp(-u)  \\
   &=&  \dfrac{T}{(1+T)^{m+1}} \exp \left( \delta_{\mathrm{max}} \right)  \int ^ { v_\mathrm{max}}  _{0 }  v^m \exp( - v) dv,\nonumber
   \end{eqnarray}
where 
$ v_\mathrm{max} = \lambda - \dfrac{T + 1}{T} \delta_{\mathrm{crit}}$.
To calculate $I_m$ we use 
\begin{align}
   J _m &=   \int ^ a _0  v^m \exp( - v) dv = m J_{m-1}  - a^m \exp(-a), \\
   J_0 &= 1 - \exp(-a).\nonumber
\end{align} 
This leads to 
\begin{align}
\label{intdPdR}
   &\int _{\delta_{\mathrm{crit}}} ^{\infty }  \exp(\delta)
   \dfrac{d {\cal P} _R}{d R}(\delta) d \delta =
  \int _{\delta_{\mathrm{crit}}} ^{\delta _{\mathrm{max}} }  \exp(\delta) 
   \dfrac{d {\cal P} _R}{d R}(\delta) d \delta = \\
   &\dfrac{e ^{-\lambda}}{T}   \dfrac{d \lambda}{dR}   \sum _{m=1} 
\left(    \left( m  - \dfrac{\lambda  (2 + T)}{1 + T } \right) I_{m-1}  +  \dfrac{\lambda  (m - 1)}{1 + T} I_{m-2} \right)     
     \dfrac{\lambda ^{m-1}}{(m-1) !  m !} , \nonumber
\end{align} 
and 
\begin{equation}
    \dfrac{d \lambda}{d R} = - \dfrac{\sigma_0^2}{2 T^2} (n_d -3) \dfrac{R ^{n_d-4}}{L_i ^{n_d-3}} .
\end{equation}

Thus, with Eqs.~(\ref{crit_Mtot_norm},\ref{intdPdR}), the mass spectrum is obtained 
from Eq.~(\ref{n_general}).
To calculate Eq.~(\ref{intdPdR}), it is convenient to calculate 
\begin{align}
K_m &= \dfrac{\lambda^m}{m !  (m+1) !}   I _m\\  
 \nonumber
&=\dfrac{\lambda^m}{m !  (m+1) !}   \int _{\delta_{\mathrm{crit}}} ^{\delta _{\mathrm{max}}  } d\delta  \exp(\delta) u^m \exp(-u) \\
 \nonumber
&= \dfrac{\lambda^m}{m !  (m+1) !}    \dfrac{T}{(1+T)^{m+1}} \exp \left( \delta_{\mathrm{max}} \right)  \int ^ {v_\mathrm{max}}  _{0 }  v^m \exp( - v) dv \\
      \nonumber
&=   \dfrac{\lambda^m}{m !  (m+1) !}    \dfrac{T}{(1+T)^{m+1}} \exp \left(  \delta_{\mathrm{max}}\right)  \left( m J_{m-1}  - a^m \exp(-a) \right) \\ 
 \nonumber
&=  \dfrac{\lambda}{1+T} \dfrac{K_{m-1}}{m+1} - H_{m},
   \end{align} 
and 
\begin{align}
H_{m} &=  \dfrac{a^ m \lambda^m}{m !  (m+1) !}    \dfrac{T}{(1+T)^{m+1}} \exp \left(  \dfrac{\lambda T}{1 + T} - a \right) 
\\&=  \dfrac{a \lambda}{m   (m+1) (1+T)} H_{m-1}  . \nonumber
 \end{align} 
Thus we get
\begin{align}
   &\int _{\delta_{\mathrm{crit}}} ^{\infty }  \exp(\delta)
   \dfrac{d {\cal P} _R}{d R}(\delta) d \delta = \\
   &\quad \dfrac{e ^{-\lambda}}{T}   \dfrac{d \lambda}{dR}   \sum _{m=1} 
\left(    \left( m  - \dfrac{\lambda  (2 + T)}{1 + T } \right) K_{m-1}  +  \dfrac{\lambda^2}{m (1 + T)} K_{m-2} \right). \nonumber
  \label{intdPdR2}
\end{align}

\section{The impact of the cut-off term on the SFR}
\label{cut-off}

\begin{figure}
\centering
\includegraphics[width=\linewidth]{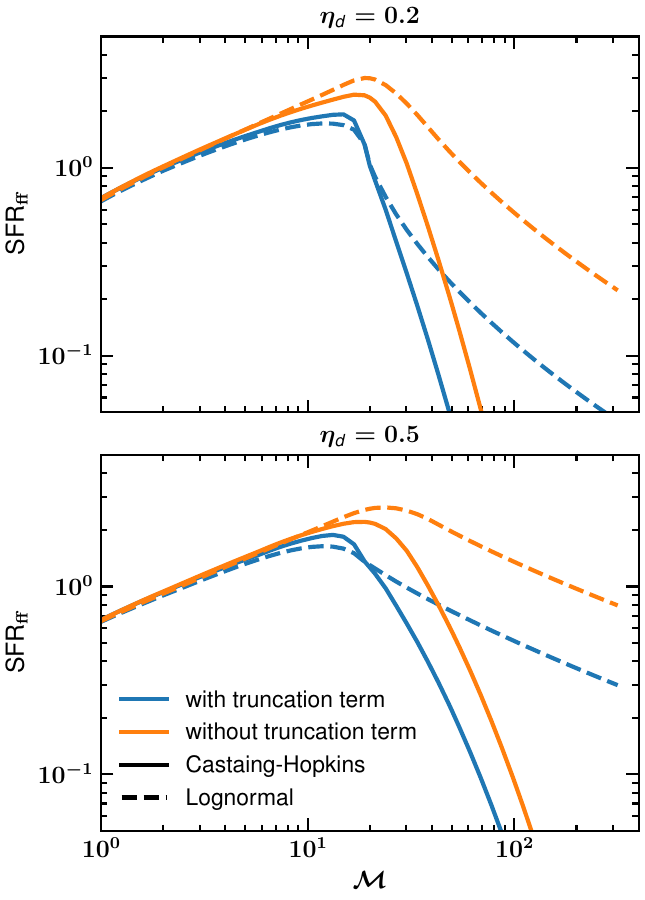}  
\caption{Star formation rate as a function of Mach number. Dashed-lines correspond to the log-normal PDF
and solid lines to the Castaing-Hopkins PDF. Blue lines stand for the complete expression of the SFR whereas 
orange lines represent the SFR obtained without the truncation term (see Eq.~\ref{n_general}). The model parameters are 
$L=200$ pc, $b=0.67$, $n_0=1.5$ cm$^{-3}$ and $\eta_d = 0.2$ (top) or $\eta_d = 0.5$ (bottom). }
\label{sfr_mass_trunc}
\end{figure}

As discussed in  Sect.~\ref{fig:mass_spectrum}, the mass spectrum is composed of two terms. The second, which entails the 
 derivative of the density PDF with respect to the spatial scale, is important only for spatial scales close to the injection length and 
  it is playing a role in reducing the SFR. To explicitly study its influence, Fig.~\ref{sfr_mass_trunc} 
 displays the SFR as a function of Mach number taking into account and 
 excluding the cut-off term. Both the log-normal  and the Castaing-Hopkins PDF are 
 displayed. It appears that the cut-off term has a quantitative influence on the results but qualitatively the results are similar
 when it is accounted for and when it is not. 

\end{document}